\documentclass[twocolumn,showpacs,prb]{revtex4}

\usepackage{graphicx}
\usepackage{dcolumn}
\newcommand{\be}{\begin{equation}}
\newcommand{\ee}{\end{equation}}
\newcommand{\bea}{\begin{eqnarray}}
\newcommand{\eea}{\end{eqnarray}}

\begin{document}
\title{Structural, Thermal, Magnetic and Electronic Transport Properties of the LaNi$_2$(Ge$_{1-x}$P$_x$)$_2$ System}
\author {R. J. Goetsch, V. K. Anand, Abhishek Pandey, and D. C. Johnston}
\affiliation {Ames Laboratory and Department of Physics and Astronomy, Iowa State University, Ames, Iowa 50011}

%,viewport= 20 20 160 230,clip

\date{December 7, 2011}

\begin{abstract}

Polycrystalline samples of LaNi$_2$(Ge$_{1-x}$P$_x$)$_2$ ($x=0,\ 0.25,\ 0.50,\ 0.75,\ 1$) were synthesized and their properties investigated by x-ray diffraction (XRD) measurements at room temperature and by heat capacity $C_{\rm{p}}$, magnetic susceptibility $\chi$, and electrical resistivity $\rho$  measurements versus temperature $T$ from 1.8 to 350~K\@. Rietveld refinements of powder XRD patterns confirm that these compounds crystallize in the body-centered-tetragonal ThCr$_2$Si$_2$-type structure (space group \emph{I}4/\emph{mmm}) with composition-dependent lattice parameters that slightly deviate from Vegard's Law. The $\rho(T)$ measurements showed a positive temperature coefficient for all samples from 1.8~K to 300~K, indicating that all compositions in this system are metallic.  The low-$T$ $C_{\rm p}$ measurements yield a rather large Sommerfeld electronic specific heat coefficient $\gamma = 12.4(2)$~mJ/mol\,K$^2$ for $x = 0$ reflecting a large density of states at the Fermi energy that is comparable with the largest values found for the $A$Fe$_2$As$_2$ class of materials with the same crystal structure.  The $\gamma$ decreases approximately linearly with $x$ to 7.4(1)~mJ/mol\,K$^2$ for $x=1$.  The $\chi$ measurements show nearly temperature-independent paramagnetic behavior across the entire range of compositions except for LaNi$_2$Ge$_2$, where a broad peak is observed at $\approx 300$~K from $\chi(T)$ measurements up to 1000 K that may arise from short-range antiferromagnetic correlations in a quasi-two-dimensional magnetic system.  High-accuracy Pad\'e approximants representing the Debye lattice heat capacity and Bloch-Gr\"uneisen electron-phonon resistivity functions versus $T$ are presented and are used to analyze our experimental $C_{\rm{p}}(T)$ and $\rho(T)$ data, respectively, for $1.8~{\rm K} \leq T \leq 300$~K\@.  The $T$-dependences of $\rho$ for all samples are well-described over this $T$ range by the Bloch-Gr\"uneisen model, although the observed $\rho(300$~K) values are larger than calculated from this model.  A significant $T$-dependence of the Debye temperature determined from the $C_{\rm p}(T)$ data was observed for each composition.  No clear evidence for bulk superconductivity or any other long-range phase transition was found for any of the LaNi$_2$(Ge$_{1-x}$P$_x$)$_2$ compositions studied.

\end{abstract}


\maketitle

\section{Introduction}

The search for high temperature superconductors intensified after the discovery of superconductivity at 26~K in the compound LaFeAsO$_{1-x}$F$_x$.\cite{Kamihara2008, Johnston2010} Subsequent studies revealed even higher superconducting transition temperatures ($T_{\rm{c}}$) upon replacing La with smaller rare earth elements,\cite{Kito2008, Ren2008b} yielding a $T_{\rm{c}}$ of 55~K for SmFeAsO$_{1-x}$F$_x$.\cite{Ren2008a} These compounds crystallize in the primitive tetragonal ZrCuSiAs (1111-type) structure with space group \emph{P}4/\emph{nmm}.\cite{Quebe2000} They have alternating FeAs and {\it R}O ({\it R} $=$ rare earth element) layers stacked along the $c$-axis. The Fe atoms form a square lattice in the $ab$~plane and are coordinated by As tetrahedra, where the coordinating As atoms lie in planes on either side of and equidistant from an Fe plane.  The undoped parent compounds show coupled structural and antiferromagnetic (AF) spin density wave (SDW) transitions,\cite{GChen2008a, Dong2008, Klauss2008} which are both suppressed upon doping by partially substituting O by F\@. Such doping results in a nonintegral formal oxidation state for the Fe atoms.  This suppression of the long-range ordering transitions appears necessary for the appearance of high-temperature superconductivity, as in the layered cuprate high-$T_{\rm c}$ superconductors.\cite{XChen2008, GChen2008a, Ren2008a, Yang2008, Bos2008, Dong2008, Giovannetti2008}  The 1111-type self-doped Ni-P analogue LaNiPO becomes superconducting at temperatures up to about 4.5~K.\cite{Tegel2008a, McQueen2009}

Subsequently, the parent compounds {\it A}Fe$_2$As$_2$ ({\it A} $=$ Ca, Sr, Ba, and Eu) were investigated.\cite{Johnston2010} They crystallize in the body-centered-tetragonal ThCr$_2$Si$_2$ (122-type) structure  with space group \emph{I}4/\emph{mmm} and contain the same type of FeAs layers as in the 1111-type compounds.  In addition, they show SDW and structural transitions at high temperatures\cite{Rotter2008a, Krellner2008, Ni2008a, Yan2008, Ni2008b, Ronning2008, Goldman2008, Tegel2008, Ren2008c, Jeevan2008a} that are similar to those seen in the 1111-type compounds and these are similarly suppressed upon substituting on the {\it A}, Fe and/or As sites.\cite{Johnston2010} Superconductivity again appears at temperatures up to 38~K as these long-range crystallographic and magnetic ordering transitions are suppressed.\cite{Rotter2008b, GChen2008b, Jeevan2008b, Sasmal2008}  However, the conventional electron-phonon interaction has been calculated to be insufficient to lead to the observed high $T_{\rm c}$'s, and strong AF fluctuations still occur in these compounds above $T_{\rm c}$ even after the long-range AF ordering is suppressed.\cite{Johnston2010}  The consensus is therefore that the superconductivity in these high-$T_{\rm c}$ compounds arises from an electronic/magnetic mechanism rather than from the conventional electron-phonon interaction.\cite{Johnston2010}

The above discoveries motivated studies of other compounds with the 122-type structure to search for new superconductors and to clarify the materials features necessary for high-$T_{\rm c}$ superconductivity in the {\it A}Fe$_2$As$_2$-type compounds.  For example, the semiconducting AF compound ${\rm BaMn_2As_2}$ contains local Mn magnetic moments with spin $S = 5/2$ and N\'eel temperature $T_{\rm N} = 625$~K.\cite{Singh2009a, An2009, Singh2009b, Johnston2011}  We recently doped this compound with K to form a new series of metallic AF Ba$_{1-x}$K$_x$Mn$_2$As$_2$ compounds containing the Mn local magnetic moments,\cite{Pandey2012} but no superconductivity has yet been observed in this system.\cite{Pandey2012, Bao2012}  This may be because the $T_{\rm N}$ was not sufficiently suppressed by the K-doping levels used.  Studies of 122-type compounds in which the Fe and As in {\it A}Fe$_2$As$_2$ are both completely replaced by other elements have also been carried out.  For example, ${\rm LaRu_2P_2}$ becomes superconducting at $T_{\rm c} = 4.1$~K.\cite{Jeitschko1987a}  ${\rm SrPd_2Ge_2}$ was recently found to become superconducting with $T_{\rm c} = 3.0$~K,\cite{Fujii2009} with conventional electronic and superconducting properties.\cite{Kim2012}

The Fe-based phosphides not containing magnetic rare earth elements such as CaFe$_2$P$_2$,\cite{Jia2010} LaFe$_2$P$_2$,\cite{Morsen1988} SrFe$_2$P$_2$,\cite{Morsen1988} and BaFe$_2$P$_2$ (Ref.~\onlinecite{Arnold2011}) show Pauli paramagnetic behavior. The 122-type Co-based phosphides exhibit varying magnetic properties.  SrCo$_2$P$_2$ does not order magnetically, although it has a large Pauli susceptibility $\chi\sim 2\times10^{-3}~{\rm cm^3/mol}$ that has variously been reported to exhibit either a weak broad peak at $\sim 110$~K attributed to ``weak exchange interactions between itinerant electrons'' (Ref.~\onlinecite{Morsen1988}), or a weak broad peak at $\sim 20$~K attributed to a ``nearly ferromagnetic Fermi liquid'' (Ref.~\onlinecite{Jia2009}).  LaCo$_2$P$_2$ orders ferromagnetically at a Curie temperature $T_{\rm C} \approx 130$~K,\cite{Morsen1988, Kovnir2011} and CaCo$_2$P$_2$ is reported to exhibit A-type antiferromagnetism at $T_{\rm N} = 113$~K in which the Co spins align ferromagnetically within the basal plane and antiferromagnetically along the {\it c}-axis.\cite{Reehuis1998}  These differing magnetic properties of the Co-based phosphides are correlated with the formal oxidation state of the Co atoms, taking into account possible P-P bonding.\cite{Reehuis1998}  Compounds where the Co atoms have a formal oxidation state of +2, $\sim +1.5$, and $\lesssim +1$ show no magnetic order, ferromagnetic order and antiferromagnetic order, respectively.\cite{Reehuis1998}  None of the above 122-type Fe or Co phosphides were reported to become superconducting.

Among Ni-containing 122-type compounds, superconductivity has been reported with $T_{\rm c} = 0.62$~K in ${\rm SrNi_2As_2}$,\cite{Bauer2008} $T_{\rm c} = 0.70$~K in the distorted structure of ${\rm BaNi_2As_2}$,\cite{Ronning2008d} $T_{\rm c} = 1.4$~K in the orthorhombically distorted structure of ${\rm SrNi_2P_2}$ (Ref.~\onlinecite{Ronning2009}) and $T_{\rm c} = 3.0$~K in ${\rm BaNi_2P_2}$.\cite{Mine2008}  The Pauli paramagnet ${\rm LaNi_2P_2}$ is reported not to become superconducting above 1.8~K.\cite{Jeitschko1987} There are conflicting reports about the occurrence of superconductivity in ${\rm LaNi_2Ge_2}$ with either $T_{\rm{c}}=0.69$--0.8~K,\cite{Wernick1982, Maezawa1999} or no  superconductivity observed above 0.32~K.\cite{Kasahara2008}

Several studies have been reported on the normal state properties of ${\rm LaNi_2Ge_2}$.  de Haas van Alphen (dHvA) measurements at 0.5~K indicated moderate band effective masses $m^*/m_{\rm e} = 1.2$ to 2.7, where $m_{\rm e}$ is the free electron mass.\cite{Maezawa1999}  Electronic structure calculations were subsequently carried out by Yamagami using the all-electron relativistic linearized augmented plane wave method based on the density-functional theory in the local-density approximation.\cite{Yamagami1999}  The density of states at the Fermi energy $E_{\rm F}$ was found to be large, ${\cal D}(E_{\rm F}) = 5.38$~states/(eV\,f.u.)\ for both spin directions, arising mainly from the Ni $3d$ orbitals, where f.u.\ means formula unit.  This ${\cal D}(E_{\rm F})$ is comparable to the largest values reported for the FeAs-based 122-type superconductors and parent compounds.\cite{Johnston2010}  Three bands were found to cross $E_{\rm F}$, with two Fermi surfaces that were hole-like (0.16 and 1.11 holes/f.u.) and one that was electron-like (0.27 electrons/f.u.), and therefore with a net uncompensated carrier charge density of 1.00 holes/f.u.  Thus for the hypothetical compound ${\rm ThNi_2Ge_2}$ one can assign formal oxidation states Th$^{+4}$, Ni$^{+2}$ and Ge$^{-4}$.  Then substituting trivalent La for tetravalent Th yields a net charge carrier concentration of one hole per formula unit.  The electron Fermi surface is a slightly corrugated cylinder along the $c$-axis centered at the X~point of the Brillouin zone, indicating quasi-two-dimensional character, similar to the electron Fermi surface pockets in the FeAs-based 122-type compounds.\cite{Johnston2010}  On the other hand, the two hole Fermi surfaces are three-dimensional and are centered at the X and Z (or M, depending on the definition\cite{Johnston2010}) points of the Brillouin zone.  These calculated Fermi surfaces were found to satisfactorily explain the results of the above dHvA measurements,\cite{Maezawa1999} including the measured band masses.  From a comparison of the calculated ${\cal D}(E_{\rm F})$ with that obtained from experimental electronic specific heat data, Yamagami inferred that many-body enhancements of the theoretical band masses are small.\cite{Yamagami1999}

Hall effect measurements on single crystals of ${\rm LaNi_2Ge_2}$ are consistent with the occurrence of multiple electron and hole Fermi surfaces, with the weakly $T$-dependent Hall coefficients given by a positive (hole-like) value $R_{\rm H} \sim +3 \times 10^{-10}$~m$^3$/C for the applied magnetic field ${\bf H}$ parallel to the $a$-axis and a negative (electron-like) value $R_{\rm H} \sim -2 \times 10^{-10}$~m$^3$/C for ${\bf H}$ parallel to the $c$-axis.\cite{Sato1998}   The thermoelectric power obtained on a polycrystalline sample of ${\rm LaNi_2Ge_2}$ is negative.\cite{Schneider1983}  

Herein we report our results on the mixed system LaNi$_{2}$(Ge$_{1-x}$P$_x$)$_{2}$.  For $x = 0$ or~1, alternating La and NiGe or NiP layers, respectively, are stacked along the $c$-axis.  We wanted to investigate whether any new phonomena occur with Ge/P mixtures that do not occur at the endpoint compositions, such as happens when the parent FeAs-based compounds are doped/substituted to form high-$T_{\rm c}$ superconductors.  In addition, Yamagami's electronic structure calculations for ${\rm LaNi_2Ge_2}$ discussed above\cite{Yamagami1999} indicated some similarities to the electronic structures of the FeAs-based 122-type compounds.

We report structural studies using powder x-ray diffraction (XRD) measurements at room temperature, together with heat capacity $C_{\rm p}$, magnetic susceptibility $\chi$, and electrical resistivity $\rho$ measurements versus temperature $T$ from 1.8 to 350~K for five compositions of LaNi$_{2}$(Ge$_{1-x}$P$_x$)$_{2}$ with $0\leq x\leq1$. Our low-$T$ limit of 1.8~K precluded checking for superconductivity with $T_{\rm c}< 1$~K reported for ${\rm LaNi_2Ge_2}$,\cite{Wernick1982, Maezawa1999} but we did find evidence for the onset of superconductivity below $\sim 2$~K in two samples of ${\rm LaNi_2P_2}$ from both $\rho(T)$ and $\chi(T)$ measurements.  However, it is not clear from our measurements whether this onset arises from the onset of bulk superconductivity or is due to an impurity phase.

Also presented in this paper is the construction of Pad\'e approximants\cite{Pade} for the Debye and Bloch-Gr\"uneisen functions that describe the acoustic lattice vibration contribution to the heat capacity at constant volume $C_{\rm V}(T)$ of materials and the contribution to the $\rho(T)$ of metals from scattering of conduction electrons from acoustic lattice vibrations, respectively. These Pad\'e approximants were created in order to easily fit our respective experimental data using the method of least-squares, but they are of course more generally applicable to fitting the corresponding data for other materials.  The Debye and Bloch-Gr\"uneisen functions themselves cannot be easily used for nonlinear least-squares fits to experimental data because they contain integrals that must be evaluated numerically at the temperature of each data point for each iteration. Several numerical expressions representing the Bloch-Gr\"uneisen\cite{Deutsch1987, Mamedov2007, Ansari2010, Cvijovic2011, Paszkowski1999} or Debye\cite{Ng1970} functions have appeared. However, they replace the integrals in these functions with infinite series, use very large numbers of terms, and/or use special functions.  These approximations are therefore not widely used for fitting experimental data.  One paper presented a method for approximating the Debye function using the Einstein model.\cite{Listerman1979} This method is also of little use for fitting because it uses a different equation for each temperature range and it becomes inaccurate at low temperatures.  However, as we demonstrate, the Debye and Bloch-Gr\"uneisen functions  can each be accurately approximated by a simple Pad\'e approximant over the entire $T$ range. To our knowledge, there are no previously reported Pad\'e approximants for either of these two important  functions.  The $T$-dependences of $\rho$ for all samples discussed here are well-described by the Bloch-Gr\"uneisen prediction, although the observed $\rho(300$~K) values are larger than calculated.  A significant $T$-dependence of the Debye temperature determined from the $C_{\rm p}(T)$ data was observed for each composition.  

The remainder of this paper is organized as follows. An overview of the experimental procedures and apparatus used in this work is given in Sec.~\ref{ExpDetails}. The construction of the Pad\'e approximants for the Bloch-Gr\"uneisen and Debye functions is described in Sec.~\ref{Functions} and Appendix~\ref{Pade}. The structural, thermal, magnetic, and electrical resistivity measurements of the LaNi$_{2}$(Ge$_{1-x}$P$_x$)$_{2}$ system and their analyses are presented in Sec.~\ref{Results} and Appendices~\ref{RietveldPlots} and~\ref{Sec:MHPlots}.  A summary and our conclusions are given in Sec.~\ref{Conclusion}.

\section{\label{ExpDetails} EXPERIMENTAL DETAILS}

Polycrystalline samples of LaNi$_{2}$(Ge$_{1-x}$P$_x$)$_{2}$ ($x$ = 0, 0.25, 0.50, 0.75, 1) were prepared using the high purity elements Ni: 99.9+\%, P: 99.999+\%, and Ge: 99.9999+\% from Alfa Aesar and La: 99.99\% from Ames Laboratory Materials Preparation Center. Stoichiometric amounts of La, Ni, and Ge were first melted together using an arc furnace under high-purity Ar atmosphere. The arc-melted button was flipped and remelted five times to ensure homogeneity. Next, the samples (except for LaNi$_{2}$Ge$_2$ which was prepared following the general procedures outlined in Ref.~\onlinecite{Rieger1969}) were throughly ground and mixed with the necessary amount of P powder in a glove box under an atmosphere of ultra high purity He. The powders were cold-pressed into pellets and placed in 2~mL alumina crucibles.  The arc-melted button of LaNi$_2$Ge$_2$ was wrapped in Ta foil.  The samples were then sealed in evacuated quartz tubes and fired at 990~$^\circ$C for $\approx 6$ d. Samples containing phosphorus were first heated to 400~$^\circ$C to prereact the phosphorus. 

After the first firing, the phase purities of the samples were checked using room temperature powder x-ray diffraction (XRD) with a Rigaku Geigerflex powder diffractometer and CuK$_\alpha$ radiation. The x-ray patterns were analyzed for impurities using {\tt MDI Jade 7}. If necessary, samples were thoroughly reground and repelletized (except for LaNi$_{2}$Ge$_{2}$ which was just rewrapped in Ta foil) and either placed back in alumina crucibles or wrapped in Ta foil and resealed in evacuated quartz tubes. Samples were again fired at 990 $^{\circ}$C for 5--6 d. The LaNi$_2$P$_2$ sample was arc-melted with additional La and P in order to achieve a single phase sample.  This may have been necessary because the compound may not form with the exact 1:2:2 stoichiometry.  After arc-melting, part of the sample was annealed for 60~h at 1000~$^{\circ}$C\@. Throughout this paper, the annealed LaNi$_2$P$_2$ sample will be referred to as $x=1.00$a and the as-cast sample as $x=1.00$b where $x$ is the composition of LaNi$_2$(Ge$_{1-x}$P$_x$)$_2$. As seen in the XRD patterns and fits in Sec.~\ref{Structure}, all final samples were single-phase except for two samples showing very small concentrations of impurities.

Rietveld refinements of the XRD patterns were carried out using the {\tt FullProf} package.\cite{Rodriguez1993}  Magnetization measurements versus applied magnetic field $H$ and temperature $T$ were carried out using a superconducting quantum interference device (SQUID) magnetometer (Quantum Design, Inc.).  Gel caps were used as sample holders and their diamagnetic contribution was measured separately and corrected for in the data presented here.

The $C_{\rm p}(T)$ and $\rho(T)$ measurements were carried out using a Quantum Design Physical Property Measurement System (PPMS). Samples for heat capacity measurements had masses of 15--40~mg and were attached to the heat capacity puck with Apiezon~N grease for thermal coupling to the platform.  The $\rho(T)$ measurements utilized a four-probe ac technique using the ac transport option on the PPMS\@.  Rectangular samples were cut from the sintered pellets or arc-melted buttons using a jeweler's saw.  Platinum leads were attached to the samples using EPO-TEK P1011 silver epoxy.  The sample was attached to the resistivity puck with GE~7031 varnish. Temperature-dependent $\rho$ measurements were recorded on both cooling and heating to check for thermal hysteresis. No significant hysteresis was observed for any of the samples. In addition, the vibrating sample magnetometer (VSM) option on the PPMS was used to measure the high-$T$ magnetization of the LaNi$_2$Ge$_2$ sample up to 1000~K\@.
 
\section{\label{Functions} Pad\'e Approximant Fits to the Bloch-Gr\"uneisen and Debye Functions}

A Pad\'e approximant $g(T)$ is a ratio of two polynomials. Here we write these polynomials as series in $1/T$ according to
\be
g(T) = \frac{N_0+\frac{N_1}{T}+\frac{N_2}{T^2}+\cdots}{D_0+\frac{D_1}{T}+\frac{D_2}{T^2}+\cdots}.
\ee
The first one, two or three and last one, two or three in each of the sets of coefficients $N_i$ and $D_i$ in $g(T)$ can be chosen to exactly reproduce \emph{both} the low- and high-$T$ limiting values and power law dependences in $T$ and/or $1/T$ of the function it is approximating.  This is a very important and powerful feature of the Pad\'e approximant.  Then the remaining terms in powers of 1/$T$ in the numerator and denominator have freely adjustable coefficients that are chosen to fit the intermediate temperature range of the function.  A physically valid approximant requires that there are no poles of the approximant on the positive real $T$ axis.

\subsection{\label{Gruneisen} Bloch-Gr\"uneisen Model}

The temperature-dependent electrical resistivity due to scattering of conduction electrons by acoustic lattice vibrations in monatomic metals is described by the Bloch-Gr\"uneisen (BG) model according to\cite{Blatt1968}
\be
 \rho(T) = 4 \mathcal{R}(\Theta_{\rm R}) \left( \frac{T}{\Theta _{\rm{R}}}\right) ^5 \int_0^{\Theta_{\rm R}/T}{\frac{x^5}{(e^x-1)(1-e^{-x})}dx}
\label{eq:Gruneisen}
\ee
where
\be
\mathcal{R}(\Theta_{\rm R}) =\frac{\hbar}{e^2} \left[ \frac{\pi^3 (3 \pi^2)^{1/3} \hbar^2}{4 n_{\rm{cell}}^{2/3} a M k_{\rm{B}} \Theta_{\rm{R}}} \right],
 \label{eq:R}
\ee
$\Theta_{\rm R}$ is the Debye temperature determined from resistivity measurements, $\hbar$ is Planck's constant divided by $2\pi$, $n_{\rm{cell}}$ is the number of conduction electrons per atom, $M=(\text{atomic weight})/N_{\rm{A}}$ is the atomic mass, $N_{\rm A}$ is Avogadro's number, $a=(\text{volume/atom})^{1/3}$, $k_{\rm{B}}$ is Boltzmann's constant, and $e$ is the elementary charge. These variables map a monatomic metal with arbitrary crystal structure onto a simple-cubic lattice with one atom per unit cell of lattice parameter $a$.  To calculate $\mathcal{R}(\Theta_{\rm R})$ in units of $\Omega$\,cm, one sets the prefactor ($\hbar/e^2$) in Eq.~(\ref{eq:R}) to 4108.24 $\Omega$ in SI units and calculates the quantities inside the square brackets in cgs units so that the quantity in square brackets has net units of cm.  If one instead has a polyatomic solid, one can map the parameters of that solid onto those of the monatomic solid described by the Bloch-Gr\"uneisen model as explained in Sec.~\ref{Transport} below.

\begin{figure}
	\includegraphics[width=3.3in]{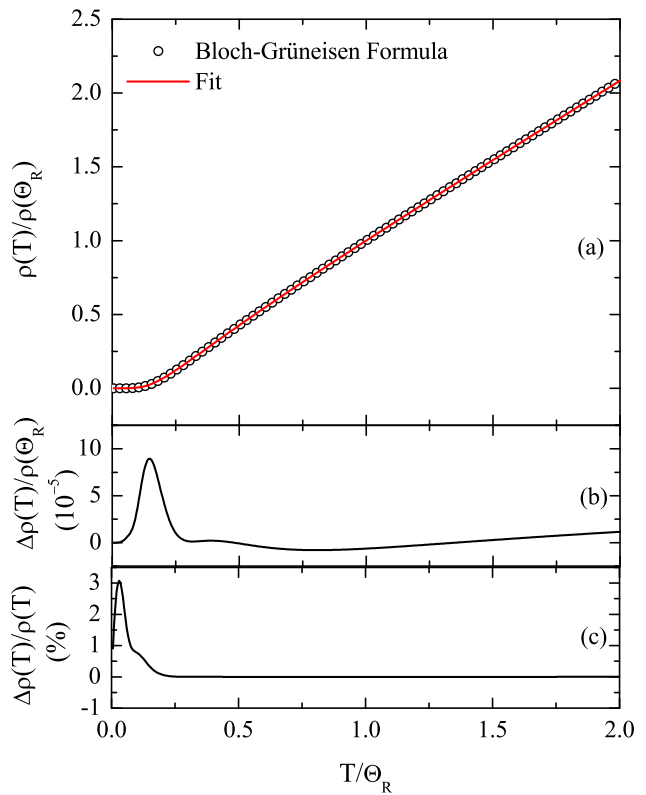}
	\caption{(Color online) (a) Normalized Bloch-Gr\"uneisen electrical resistivity in Eq.~(\ref{eq:norm_Gruneisen}) (one in every five points used for fitting are plotted, open circles) and the Pad\'e approximant fit (solid red curve), (b) Residuals (Pad\'e approximant value minus Bloch-Gr\"uneisen formula value), and (c) Percent error (residual divided by value of the Bloch-Gr\"uneisen formula), all versus temperature $T$ divided by the Debye temperature $\Theta_{\rm R}$.}
	\label{fig:Gruneisen_Fit}
\end{figure}

In practice, one fits the $T$ dependence of an experimental $\rho(T)$ data set by the BG model using an independently adjustable prefactor $\rho(\Theta_R)$ instead of $4\mathcal{R}(\Theta_{\rm R})$ in Eq.~(\ref{eq:Gruneisen}), because accurately fitting both the magnitude and $T$ dependence of a data set cannot usually be done using a single adjustable parameter $\Theta_{\rm R}$. One therefore normalizes Eq.~(\ref{eq:Gruneisen}) by $\rho(T=\Theta _{\rm{R}})$. When $T=\Theta_{\rm{R}}$, the integral in Eq.~(\ref{eq:Gruneisen}) is
\[
 \int _0 ^1{\frac{x^5}{(e^x-1)(1-e^{-x})}dx} = 0.2\,366\,159,
\]
yielding
\begin{equation}
\rho(\Theta_R)=0.9\,464\,635\, \mathcal{R}(\Theta_{\rm R}).
\label{eq:norm_rho_debye_temp}
\end{equation}
Equations~(\ref{eq:Gruneisen})--(\ref{eq:norm_rho_debye_temp}) then yield the normalized $T$-dependence of the BG function~(\ref{eq:Gruneisen}) as
\bea
\frac{\rho(T)}{\rho (\Theta _{\rm{R}})}&= & 4.226\,259 \left( \frac	{T}{\Theta _{\rm{R}}} \right)^5 \nonumber \\* \nopagebreak
&&\times \int_0^{\Theta_R/T}{\frac{x^5}{(e^x-1)(1-e^{-x})}dx}.
\label{eq:norm_Gruneisen}
\eea
This $T$-dependence is only a function of the dimensionless normalized temperature $T/\Theta_{\rm R}$.  Therefore we define normalized $\rho$ and $T$ variables as
\bea
\rho_{\rm{n}}(T_{\rm n})&=& \frac{\rho (T_{\rm n})}{\rho(\Theta _{\rm{R}})}\label{Eq:rhonDef} \\
T_{\rm{n}}&=&\frac{T}{\Theta _{\rm{R}}}\nonumber
\eea
and Eq.~(\ref{eq:norm_Gruneisen}) becomes
\bea
\rho_{\rm{n}}(T_{\rm n}) &=& 4.226\,259\, T_{\rm n}^5 \nonumber \\*
&&\times \int_0^{1/T_{\rm n}}{\frac{x^5}{(e^x-1)(1-e^{-x})}dx}.
\label{eq:norm_Gruneisen2}
\eea
A set of $\rho_{\rm{n}}(T_{\rm n})$ data points calculated from Eq.~(\ref{eq:norm_Gruneisen2}) is plotted in Fig.~\ref{fig:Gruneisen_Fit}(a). These values were then used as a set of ``data'' to fit by a Pad\'e approximant as described next.

\begin{table}
 \caption{\label{table:Gruneisen_approx} Values of the coefficients in the Pad\'e approximant in Eq.~(\ref{eq:Gruneisen_approx}) that accurately fits the normalized Bloch-Gr\"uneisen function in Eq.~(\ref{eq:norm_Gruneisen2}).}
 \begin{ruledtabular}		
		\begin{tabular}{c D{.}{.}{20}}
		 Coefficient & {\rm Value} \\ \hline
			$N_0$ & 1083.127\,77 \\
			$N_1$ & 401.679\,91 \\
			$N_2$ & -16.787\,903\,6 \\
			$N_3$ & 3.717\,146\,28 \\
			$D_1$ & 1025.140\,90 \\
			$D_2$ & 380.175\,373 \\
			$D_3$ & 41.063\,139\,0 \\
			$D_4$ & 24.580\,952\,4 \\
			$D_5$ & 0.177\,731\,204 \\
			$D_6$ & 0.586\,502\,906 \\
			$D_7$ & -0.018\,365\,823\,3 \\
			$D_8$ & 0.007\,068\,443\,59 \\
		\end{tabular}
 \end{ruledtabular}
\end{table}

In order to construct a Pad\'e approximant function that accurately represents $\rho_{\rm{n}}(T_{\rm n})$ in Eq.~(\ref{eq:norm_Gruneisen2}), the power law $T_{\rm n}$ dependences of the latter function must be computed at high and low temperatures and the coefficients of the Pad\'e approximant adjusted so that both of these limiting $T$ dependences are exactly reproduced (to numerical precision).  Then the remaining coefficients are determined by fitting the approximant to a set of data obtained by evaluating the function over the entire relevant temperature range.  This procedure is described in Appendix~\ref{AppBG}, which constrains the values of $D_1$, $D_2$, $D_3$, and $D_8$.  The resulting approximant is
\be
\rho_{\rm{n}}(T_{\rm n}) = \frac{N_0+\frac{N_1}{T_n}+\frac{N_2}{{T_n}^2}+\frac{N_3}{{T_n}^3}}{\frac{D_1}{T_n}+\frac{D_2}{{T_n}^2}+\cdots +\frac{D_7}{{T_n}^7}+ \frac{D_8}{{T_n}^8}}.
\label{eq:Gruneisen_approx}
\ee
where the fitted coefficients in the Pad\'e approximant (\ref{eq:Gruneisen_approx}) are listed in Table~\ref{table:Gruneisen_approx}. The denominator in Eq.~(\ref{eq:Gruneisen_approx}) was checked for zeros and all were found not to lie on the positive $T_{\rm{n}}$ axis. This insures that the approximant does not diverge at any (real positive) temperature. As seen in Fig.~\ref{fig:Gruneisen_Fit}(b), the difference between the BG function and the Pad\'e approximant is less than $1\times10^{-4}$ at any $T$ in the range $0<T/\Theta_{\rm{R}}<2$. By construction, the Pad\'e approximant must asymptote to the exact BG $T$ dependences at high- and low-$T$, respectively. Figure~\ref{fig:Gruneisen_Fit}(c) shows the percent error of the fitted approximant. This error is largest at low $T$ with a value of $\approx 3\%$ at $T/\Theta_{\rm{R}} \approx 0.1$. This is acceptable considering the very small value of $\rho_{\rm n}$ at such low $T_{\rm n}$. For the most accurate fit of low-$T$ experimental data by a power law in $T$, one would directly fit experimental data by the power law rather than using the Pad\'e approximant function.

When fitting experimental $\rho(T)$ data by the Pad\'e approximant $\rho_{\rm n}(T_{\rm n})$ in Eq.~(\ref{eq:Gruneisen_approx}), one fits only the $T$ dependence and not the magnitude of $\rho(T)$ by the BG theory, because as noted above, one cannot in general obtain a good fit of both the magnitude and the $T$ dependence of a measured $\rho(T)$ data set using only the single fitting parameter $\Theta_{\rm R}$. Thus, one fits an experimental $\rho(T)$ data set by
\be
\rho(T)=\rho_0 + {\rho(\Theta _{\rm{R}})}\, \rho_{\rm n}(T/\Theta_{\rm R}),
\label{eq:Gruneisen_approx_final}
\ee
where $\rho_0$ is the residual resistivity for $T \to 0$.  The meaning of $\rho_{\rm n}(T/\Theta_{\rm R})$ is that one substitutes $T/\Theta_{\rm R}$ for $T_{\rm n}$, according to Eqs.~(\ref{Eq:rhonDef}), in the Pad\'e approximant function $\rho_{\rm n}(T_{\rm n})$ in Eq.~(\ref{eq:Gruneisen_approx}).  The three adjustable parameters $\rho_0$, $\rho(\Theta _{\rm{R}})$ and $\Theta_{\rm R}$ are varied independently to obtain a fit to the data.  Once a good fit is obtained and all three parameters are  determined, one can compare the measured value of $\rho(T = \Theta_{\rm R})$ with the value predicted by the BG theory in Eq.~(\ref{eq:norm_rho_debye_temp}).  Often the agreement is not very good even for $s$- or $sp$-metals.\cite{Blatt1968}

\subsection{\label{Debye_Function} Debye Model}

\begin{figure}
	\includegraphics[width=3.3in]{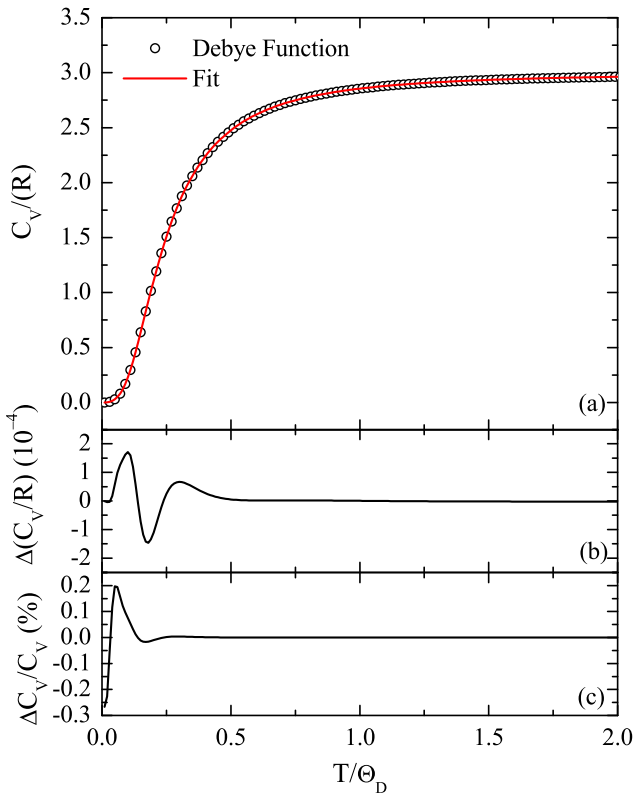}
	\caption{(Color online) (a) Plot of normalized Debye function in Eq.~(\ref{eq:norm_Debye}) (One in every two points used for fitting are plotted) (open circles) and Pad\'e approximant (red line). (b) Plot of residuals (Pad\'e approximant minus Debye function). (c) Percent error (residual divided by value of the Debye function).}
	\label{fig:Debye_Fit}
\end{figure}

The Debye model\cite{Kittel2005} is widely used for fitting experimental heat capacity $C_{\rm p}(T)$ data taken at constant pressure~p arising from acoustic lattice vibrations. It is sometimes useful to fit a large $T$ range of experimental $C_{\rm p}$ data using the Debye function.  Here we describe the construction of an accurate Pad\'e approximant function of $T$ that can easily be used in place of the Debye function~(\ref{eq:Debye}) for least-squares fitting experimental $C_{\rm p}(T)$ data over an extended $T$ range.  It can also be conveniently used to calculate the $T$ dependence of the Debye temperature from experimental lattice heat capacity data over an extended $T$ range.

The lattice heat capacity at constant volume V per mole of atoms in the Debye model is given by\cite{Kittel2005}
\be
C_{\rm{V}}(T)=9  R \left(\frac{T}{\Theta_{\rm{D}}}\right)^3 {\int_0^{\Theta_{\rm D}/T} \frac{x^4 e^x}{(e^x-1)^2}\, dx},
\label{eq:Debye}
\ee
where $R$ is the molar gas constant and $\Theta_{\rm{D}}$ is the Debye temperature determined from heat capacity measurements. The $C_{\rm{V}}$ and $T$ can be normalized to become dimensionless according to
\bea
C_{\rm{n}}(T_{\rm{n}})&=& \frac{C_{\rm{V}}(T_{\rm n})}{R},\label{Eq:CTnorm} \\
T_{\rm{n}}&=&\frac{T}{\Theta_{\rm{D}}}.\nonumber
\eea
Equation~(\ref{eq:Debye}) then becomes
\be
C_{\rm{n}}(T_{\rm n})=9 {T_{\rm{n}}}^3 \int_0^{1/T_{\rm{n}}} \frac{x^4 e^x}{(e^x-1)^2}\, dx.
\label{eq:norm_Debye}
\ee
The $C_{\rm n}$ was calculated for a representative set of $T_{\rm n}$ values using Eq.~(\ref{eq:norm_Debye}).  The resulting data are plotted as open red circles in Fig.~\ref{fig:Debye_Fit}(a).

\begin{table}
 \caption{\label{table:Debye_coeff} Values of the coefficients in the Pad\'e approximant in Eq.~(\ref{eq:Debye_approx}) that accurately fits the normalized Debye function in Eq.~(\ref{eq:norm_Debye}).}
 \begin{ruledtabular}	
		\begin{tabular}{c D{.}{.}{20}}
		 Coefficient & {\rm Value} \\ \hline
			$N_0$ & 226.684\,46 \\
			$N_1$ & 64.752\,051\,1 \\
			$N_2$ & 17.285\,710\,5 \\
			$N_3$ & 1.052\,246\,63 \\
			$N_4$ & -0.035\,843\,776\,1 \\
			$N_5$ & 0.027\,925\,482\,7 \\
			$D_0$ & 75.561\,486\,7 \\
			$D_1$ & 21.584\,017\,0 \\
			$D_2$ & 9.539\,977\,83 \\
			$D_3$ & 1.427\,243\,1 \\
			$D_4$ & 0.337\,538\,084 \\
			$D_5$ & 0.034\,609\,046\,3 \\
			$D_6$ & 0.007\,440\,025\,83 \\
			$D_7$ & -0.000\,210\,411\,972 \\
			$D_8$ & 0.000\,119\,451\,046 \\
		\end{tabular}
 \end{ruledtabular}
\end{table}

As discussed in Appendix~\ref{PadeDebye}, the Pad\'e approximant that fits both the high- and low-$T$ power law asymptotics of $C_{\rm{n}}(T_{\rm n})$ in Eq.~(\ref{eq:norm_Debye}) and has additional terms in powers of $1/T$ in the numerator and denominator to fit the intermediate $T$ range is
\be
C_{\rm{n}}(T_{\rm{n}}) = \frac{N_0+\frac{N_1}{{T_{\rm{n}}}}+\frac{N_2}{{T_{\rm{n}}}^2}+\cdots+\frac{N_5}{{T_{\rm{n}}}^5}}{D_0+\frac{D_1}{{T_{\rm{n}}}}+\frac{D_2}{{T_{\rm{n}}}^2}+\cdots +\frac{D_7}{{T_{\rm{n}}}^7}+\frac{D_8}{{T_{\rm{n}}}^8}},
\label{eq:Debye_approx}
\ee
where the coefficients are given in Table~\ref{table:Debye_coeff}. The resulting fit and error analyses are shown in Fig.~\ref{fig:Debye_Fit}. The Pad\'e approximant does not deviate from the normalized Debye function in Eq.~(\ref{eq:norm_Debye}) by more than $2\times 10^{-4}$ at any $T$ as seen in Fig.~\ref{fig:Debye_Fit}(b).  By construction, the deviation goes to zero at both low and high $T$.  The percent error in Fig.~\ref{fig:Debye_Fit}(c) has its maximum magnitude of $0.3\%$ at low $T$, and occurs because $C_{\rm n}(T_{\rm n}\to 0)\to0$ and numerical precision becomes an issue there. For another example of the high accuracy and use of this Pad\'e approximant, see Fig.~\ref{fig:Delta_Theta_vs_T} below, where the $\Theta_{\rm D}$ versus~$T$ is calculated directly for each of our samples of LaNi$_2$(Ge$_{1-x}$P$_x$)$_2$ using the Debye function in Eq.~(\ref{eq:Debye}) and compared with that found using the Pad\'e approximant in Eq.~(\ref{eq:Debye_approx}); only small differences are found.

To fit experimental $C_{\rm p}(T)$ data by the Pad\'e approximant $C_{\rm n}(T_{\rm n})$ in Eq.~(\ref{eq:Debye_approx}), one fits both the magnitude and $T$ dependence of $C_{\rm p}$ simultaneously using
\be
C_{\rm p}(T)=n R C_{\rm n}(T/\Theta_{\rm D}),
\label{eq:final_Debye_approx1}
\ee
where $n$ is the number of atoms per formula unit and $\Theta_{\rm D}$ is the only fitting parameter. Here one substitutes $T/\Theta_{\rm D}$ for $T_{\rm n}$, according to Eqs.~(\ref{Eq:CTnorm}), in the Pad\'e approximant function $C_{\rm n}(T_{\rm n})$ in Eq.~(\ref{eq:Debye_approx}).  For a metal, one can add to Eq.~(\ref{eq:final_Debye_approx1}) a linear specific heat term $\gamma T$ giving
\be
C_{\rm p}(T)=\gamma T +n R C_{\rm n}(T/\Theta_{\rm D}).
\label{eq:CpSum}
\ee
The $\gamma$ is the Sommerfeld electronic specific heat coefficient that can be experimentally determined from a prior separate fit to $C_{\rm{p}}(T)$ data at low $T$ according to\cite{Kittel2005}
\be
\frac{C_{\rm{p}}(T)}{T} =\gamma +\beta T^2
\label{eq:gamma}
\ee
as in Fig.~\ref{fig:CT_vs_T2} below, where $\beta T^3$ is the low-$T$ limit in Eq.~(\ref{eq:lowT_Debye}) of the Debye heat capacity. In Eq.~(\ref{eq:CpSum}), the only fitting parameter is $\Theta_{\rm D}$.

\section{\label{Results} Experimental Results and Analyses}

\subsection{\label{Structure} Structure and Chemical Composition Determinations}

\begin{figure}
	\includegraphics[width=3.in]{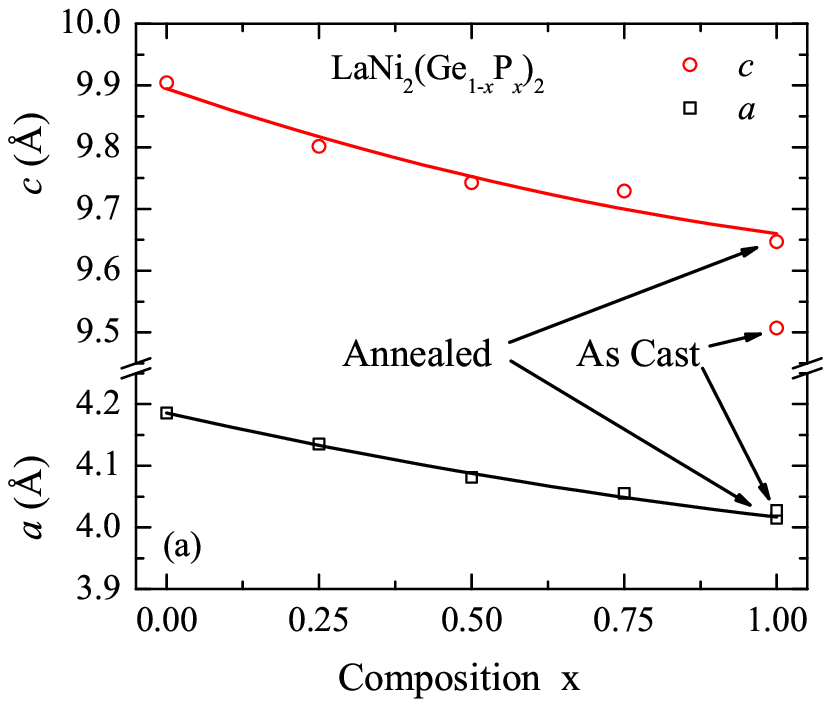}
	\includegraphics[width=3.in]{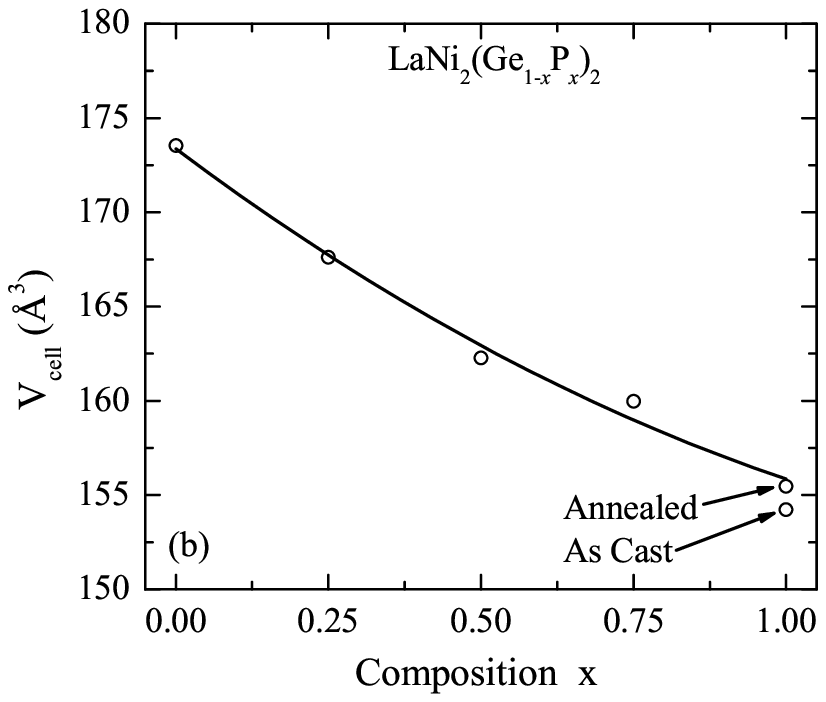}
	\caption{(Color online) (a) Unit cell lattice parameters $a$ and $c$ and (b) unit cell volume $V_{\rm cell}$ versus composition $x$ for LaNi$_2$(Ge$_{1-x}$P$_x$)$_2$. The error bars are smaller than the symbol size and the solid curves are guides to the eye.}
	\label{fig:crystallographic_comparison}
\end{figure}

\begin{figure}
	\includegraphics[width=3.3in]{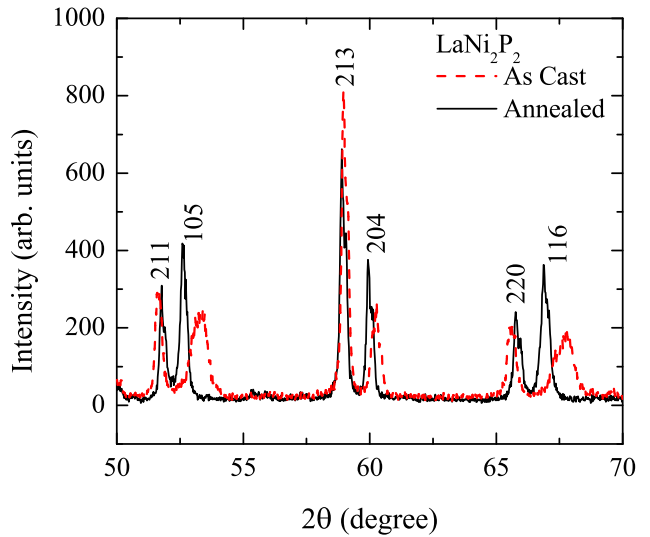}
	\caption{(Color online) Comparison of a section of the room temperature powder XRD pattern of LaNi$_2$P$_2$ before and after annealing. Numbers above the peaks are their ($hkl$) Miller indices.}
	\label{fig:LaNi2P2_comparison_XRD}
\end{figure}

\begin{table*}
 \caption{\label{table:structure} Crystallographic parameters of the body-centered-tetragonal LaNi$_2$(Ge$_{1-x}$P$_x$)$_2$ system at room temperature (space group \emph{ I}4/\emph{mmm}). Atomic coordinates are: La: (0, 0, 0), Ni: (0, 1/2, 1/4), P/Ge: (0, 0, $z_{\rm{P/Ge}}$). Listed are the  lattice parameters $a$ and $c$, the $c/a$ ratio, the unit cell volume $V_{\rm{cell}}$, and the $z$-coordinate of the P/Ge site $z_{\rm{P/Ge}}$. The quality-of-fit parameters $R_{\rm p}$, $R_{\rm wp}$ and $\chi^2$ are also listed.}
 \begin{ruledtabular}	
		\begin{tabular}{l l l l l l l l l c}
		 Compound 							& $a$(\AA) 	& $c$(\AA) 	& $c/a$ 	& $V_{\rm{cell}}$ (\AA$^3$) & $z_{\rm{P/Ge}}$ & $R_{\rm{p}}$ (\%) & $R_{\rm{wp}}$ (\%) & $\chi^2$	& Ref. \\ \hline
		LaNi$_{1.70(1)}$P$_2$				& 4.018(2)	& 9.485(6)	& 2.361		& 153.1		& 0.3716(2)	& 		&		&		& \onlinecite{Bobev2009} \\
		LaNi$_2$P$_2$ 						& 4.010(1)	& 9.604(2)	& 2.395		& 154.5		& 0.3700(2)	&		&		&		& \onlinecite{Hofmann1984} \\
											& 4.007		& 9.632		& 2.404		& 154.6		& 			&		&		&		& \onlinecite{Jeitschko1980} \\
		\ \ \ \ \ (as cast)					& 4.0276(3)	& 9.5073(9)	& 2.3605(4)	& 154.22(4) & 0.3692(7)	& 15.7	& 20.9	& 4.24	& This work	\\
		\ \ \ \ \ (annealed)				& 4.0145(2)	& 9.6471(6)	& 2.4031(3)	& 155.47(3)	& 0.3681(6)	& 13.2  & 17.7  & 3.01	& This work	\\
		LaNi$_2$(P$_{0.75}$Ge$_{0.25}$)$_2$ & 4.0550(1) & 9.7289(3) & 2.3992(1) & 159.97(1) & 0.3685(3)	& 11.5  & 16.4  & 9.05	& This work	\\
		LaNi$_2$(P$_{0.50}$Ge$_{0.50}$)$_2$ & 4.08132(7)& 9.7424(2) & 2.38707(9)& 162.281(9)& 0.3678(2)	& 10.8  & 14.4  & 3.80	& This work	\\
		LaNi$_2$(P$_{0.25}$Ge$_{0.75}$)$_2$ & 4.1353(4) & 9.8012(9) & 2.3701(4) & 167.61(5) & 0.3668(2)	& 9.40  & 13.4  & 6.35 & This work	\\
		LaNi$_2$Ge$_2$ 						& 4.18586(4)& 9.9042(1) & 2.36610(6)& 173.535(6)& 0.3678(1) & 10.2  & 13.2  & 8.66 & This work	\\
											& 4.1860(6)	& 9.902(1)	& 2.366		& 173.51	& 0.3667(2)	&		&		&		& \onlinecite{Hasegawa2004} \\
											& 4.187		& 9.918		& 			& 			&			&		&		&		& \onlinecite{Yamagami1999} \\
											& 4.187(6)	& 9.918(10)	& 2.369		& 173.8		&			&		&		&		& \onlinecite{Rieger1969} \\
											& 4.1848(2)	& 9.900(1)	& 			&			&			&		&		&		& \onlinecite{Morozkin1997}
		\end{tabular}
 \end{ruledtabular}
\end{table*}

The starting parameters for the Rietveld refinements of the powder XRD patterns were those previously reported for LaNi$_2$P$_2$ (Refs.~\onlinecite{Bobev2009}, \onlinecite{Hofmann1984}, \onlinecite{Jeitschko1980}) and LaNi$_2$Ge$_2$ (Refs.~\onlinecite{Yamagami1999}, \onlinecite{Rieger1969}, \onlinecite{Hasegawa2004}, \onlinecite{Morozkin1997}) that are presented in Table~\ref{table:structure}. All samples in the LaNi$_2$(Ge$_{1-x}$P$_x$)$_2$ system were found to crystallize in the body-centered-tetragonal ThCr$_2$Si$_2$ structure (space group \emph{I}4/\emph{mmm}) as previously reported for the compositions $x=0$ and 1, and our refined values for the lattice parameters are in agreement with reported values, as shown in Table \ref{table:structure}. The refinements of the powder XRD patterns are shown in Figs.~\ref{fig:LaNi2P1.5Ge0.5_XRD}--\ref{fig:LaNi2Ge2_XRD} in Appendix~\ref{RietveldPlots} and the crystal data are listed in Table~\ref{table:structure}. All samples were also refined for site occupancy and no significant deviations were found from the value of unity. However, the Rietveld fits were not very sensitive to changes in site occupation.

The lattice parameters $a$ and $c$ and the unit cell volume $V_{\rm cell}$ are plotted versus composition $x$ in Fig.~\ref{fig:crystallographic_comparison}.  As the concentration of P increases, the lattice parameters and unit cell volume all decrease monotonically while, from Table~\ref{table:structure}, the $c/a$ ratio increases.  The composition dependences of the quantities in Fig.~\ref{fig:crystallographic_comparison} deviate slightly from the linearities expected from Vegard's Law.  From Table~\ref{table:structure}, the $z_{\rm P/Ge}$ $c$-axis position parameter of the P/Ge site has a small overall increase with increasing P concentration. 

An interesting effect was observed in the x-ray data for LaNi$_2$P$_2$. The as-cast sample after arc-melting showed broadening of the diffraction peaks with large $c$-axis contributions.  After annealing the arc-melted sample for 60~h at 1000~$^{\circ}$C, those peaks became sharp and shifted to lower 2$\theta$ angles reflecting an increased $c$-axis lattice parameter. These effects are shown on an expanded scale in Fig.~\ref{fig:LaNi2P2_comparison_XRD} for the (105) and (116) reflections.  The $c$-axis peak broadening may arise from disorder in the interlayer stacking distances along the $c$-axis.\cite{Johnston1984}

\subsection{\label{HC} Heat Capacity Measurements}

\begin{figure}
	\includegraphics[width=3.3in]{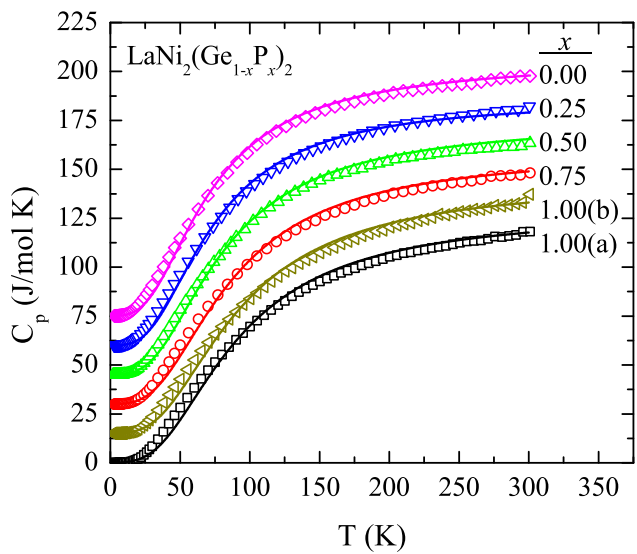}
 	\caption{(Color online) Heat capacity $C_{\rm p}$ versus temperature $T$ for samples in the LaNi$_2$(Ge$_{1-x}$P$_x$)$_2$ system (open symbols).  Fits of the data by Eq.~(\ref{eq:CpSum}), which is the sum of electronic and lattice contributions, are shown as solid curves with the respective color. For clarity, each plot is offset vertically by 15 J/mol\,K from the one below it. }
	\label{fig:Cp_vs_T}
\end{figure}

\begin{figure}
	\includegraphics[width=3.3in]{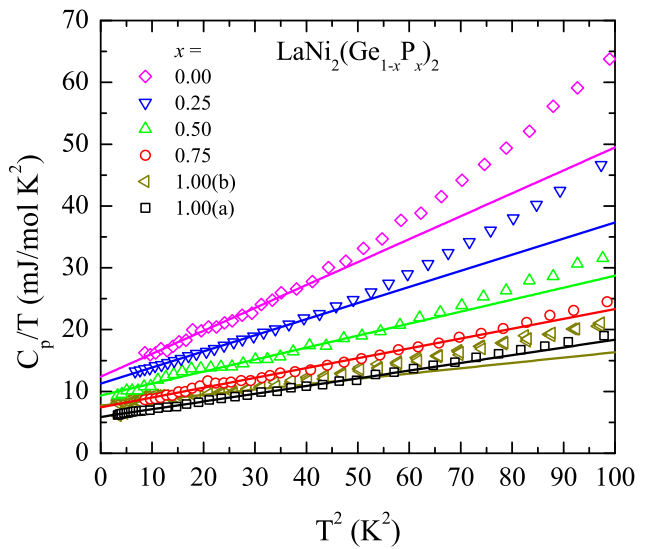}
 	\caption{(Color online) Heat capacity $C_{\rm p}$ divided by temperature $T$ versus $T^2$ (open symbols) and linear fits (solid curves of corresponding color) to the lowest-$T$ data by $C_{\rm p}/T=\gamma + \beta T^2$. Table~\ref{table:HC_properties} lists the temperature ranges of the fits and the values of $\gamma$ and $\beta$ obtained.}
	\label{fig:CT_vs_T2}
\end{figure}

\begin{figure}
	\includegraphics[width=3.3in]{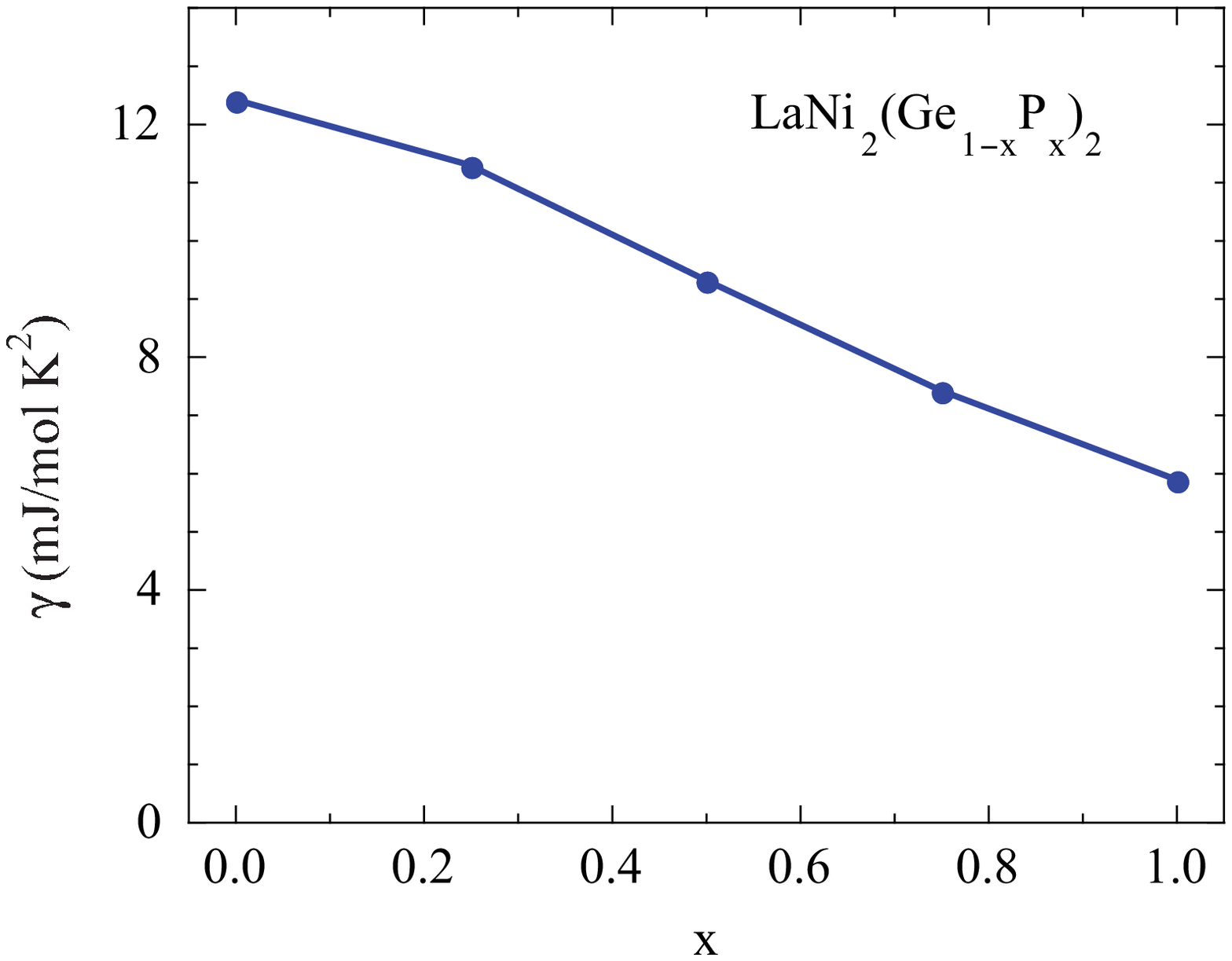}
 	\caption{(Color online) Sommerfeld electronic specific heat coefficient $\gamma$ versus composition $x$ in LaNi$_2$(Ge$_{1-x}$P$_x$)$_2$ from Table~\ref{table:HC_properties}.  For $x=1$, the datum in Table~\ref{table:HC_properties} for the annealed sample is plotted.  The error bars are smaller than the data symbols.  The line is a guide to the eye.}
	\label{Fig:gamma_vs_x}
\end{figure}

Plots of $C_{\rm p}(T)$ for our samples of LaNi$_2$(Ge$_{1-x}$P$_x$)$_2$ from 1.8 to 300~K are shown in Fig.~\ref{fig:Cp_vs_T}. The heat capacity at 300~K ranges from 118.3--123.0~J/mol\,K for these samples. These values are approaching with increasing $T$ the expected classical Dulong-Petit value $C_{\rm{V}}=3nR=124.7$~mJ/mol\,K for the heat capacity   due to acoustic lattice vibrations, where $n$ is the number of atoms per formula unit ($n=5$ for our compounds).   

To determine the Sommerfeld electronic specific heat coefficient $\gamma$ and a low temperature value of the Debye temperature $\Theta_{\rm{D}}$ for each sample, the lowest temperature linear $C_{\rm p}(T)/T$ data for each sample, plotted in Fig.~\ref{fig:CT_vs_T2}, were fitted by Eq.~(\ref{eq:gamma}) and the values of $\gamma$ and $\beta$ obtained.  A value of $\Theta_{\rm{D}}$ can be calculated from each value of $\beta$ using\cite{Kittel2005}
\be
\Theta_{\rm{D}}=\left(\frac{12\pi ^4 R n}{5 \beta} \right)^{1/3}.
\label{eq:theta}
\ee
The values obtained for $\gamma$, $\beta$, and $\Theta_{\rm D}$ for each sample are listed in Table~\ref{table:HC_properties}.  A plot of $\gamma$ versus $x$ in LaNi$_2$(Ge$_{1-x}$P$_x$)$_2$ is shown in Fig.~\ref{Fig:gamma_vs_x}, where a nearly linear decrease in $\gamma$ with increasing $x$ is seen.  Using this $\gamma$ values in Table~\ref{table:HC_properties}, we then fitted the data from 1.8 to 300~K in Fig.~\ref{fig:Cp_vs_T} by Eq.~(\ref{eq:CpSum}) and obtained the $\Theta_{\rm D}$ fitting parameters listed in Table~\ref{table:HC_properties}.  The fits are shown as the solid curves in Fig.~\ref{fig:Cp_vs_T}, which are seen to agree rather well with the respective data.  However, small $T$-dependent deviations between the data and fit for each sample are seen, which we address next.

\begin{table*}
\caption{\label{table:HC_properties} Values of $\gamma$ and $\beta$ obtained from the low-$T$ fits of the data in Fig.~\ref{fig:CT_vs_T2} by Eq.~(\ref{eq:gamma}) are listed together with the density of states at the Fermi energy ${\cal D}(E_{\rm F})$ in units of states/(eV\,f.u.) for both spin directions calculated from $\gamma$ using Eq.~(\ref{Eq:DfromGamma0}).  Also shown are the $\Theta_{\rm D}$ values calculated from the low-$T$ $\beta$  values using Eq.~(\ref{eq:theta}) and from a global fit to all the lattice $C_{\rm p}(T)$ data from 1.8 to 300~K for each sample.  Available values from the literature are also listed.}
\begin{ruledtabular}	
\begin{tabular}{l l l l l l l c}
Sample & Low-$T$ & $\gamma$ & ${\cal D}(E_{\rm F})$ & $\beta$ & $\Theta_{\rm D}$ & $\Theta_{\rm D}$ & Ref. \\ 
	& Fit Range & (mJ/mol\,K$^2$) &    (eV\,f.u.)$^{-1}$            & (mJ/mol\,K$^4$) & Low-$T$ fit & All-$T$ fit\\
	& (K) & & & & (K) & (K)\\ \hline
LaNi$_2$P$_2$   (as-cast)			& 1.81--5.34	& 7.7(2)	&   & 0.086(7)	& 483(14)	& 369(2)	& This work \\
\hspace{0.45in} (annealed)		&	1.81--7.31& 5.87(2)	& 2.49  & 0.126(2)	& 426(3)	& 365(3)	& This work\\
LaNi$_2$(P$_{0.75}$Ge$_{0.25}$)$_2$	& 2.93--8.60	& 7.4(1) 	& 3.13  & 0.159(3) 	& 394(3) 	& 348(2) 	& This work\\
LaNi$_2$(P$_{0.50}$Ge$_{0.50}$)$_2$ & 1.82--7.13	& 9.3(2) 	& 3.94  & 0.194(7) 	& 369(5)  	& 326(2) 	& This work\\
LaNi$_2$(P$_{0.25}$Ge$_{0.75}$)$_2$ & 2.59--6.52 	&11.27(6) &  4.78 & 0.261(2) 	& 333.9(9) 	& 301(2) 	& This work\\
LaNi$_2$Ge$_2$ 					& 2.93--6.41 & 12.4(2) &	5.26 & 0.371(9) 	& 297(2) 	& 287(2) 	& This work\\
								&	& 14.5 	&	&  0.273  	&   328 	& 			& \onlinecite{Kasahara2008} \\
								&	& 12.7 (calc) & 5.38&   &  	&& \onlinecite{Yamagami1999} \\
								&	& 13.5 (obs)\footnotemark[1] 	&	&	&	&	&\onlinecite{Yamagami1999}\\
\end{tabular}
\end{ruledtabular}
\footnotetext[1]{No experimental evidence or reference citation was given for this quoted observed value.}
\end{table*}

	Deviations of a Debye model fit from experimental lattice heat capacity data are due to the following assumptions and approximations of the model.
\begin{itemize}
\item{The system is assumed to be at constant volume as $T$ changes, rather than at constant pressure which is the experimental condition.  This deficiency can be corrected for if the $T$-dependent compressibility and thermal expansion coefficient are known for the compound of interest.}
\item{A quadratic density of phonon states versus energy is assumed, which terminates at the Debye energy $k_{\rm B}\Theta_{\rm D}$.  For actual materials, this assumption can only be accurately applied at temperatures $T\ll\Theta_{\rm D}$, which gives the Debye~$T^3$ law [the second term on the right-hand side of Eq.~(\ref{eq:gamma})].}
\item{Assumptions are made that the speed of acoustic sound waves in a material is temperature-independent, is isotropic and is the same for longitudinal and transverse acoustic sound waves.  In general, these assumptions are too simplistic and the parameters are temperature dependent.}
\item{The Debye model only accounts for acoustic lattice vibrations and does not take into account optic lattice vibrations arising from opposing vibrations of atoms with different masses in the unit cell.  The contribution of these to $C_{\rm p}(T)$ can be modeled by adding Einstein terms to the fit function.\cite{Kittel2005}}
\end{itemize}

	Because of these approximations and assumptions of the Debye model, the lattice heat capacity of a material is never precisely described by the Debye model over an extended temperature range such as from 2~K to 300~K\@.  The most serious approximation in our $T$ range is the second approximation.  Within the Debye model $\Theta_{\rm D}$ is independent of $T$\@.  One can therefore parameterize the deviations of a fit from the data by allowing $\Theta_{\rm D}$ to vary with $T$.\cite{Gopal1966}

\begin{figure}
	\includegraphics[width=3.3in]{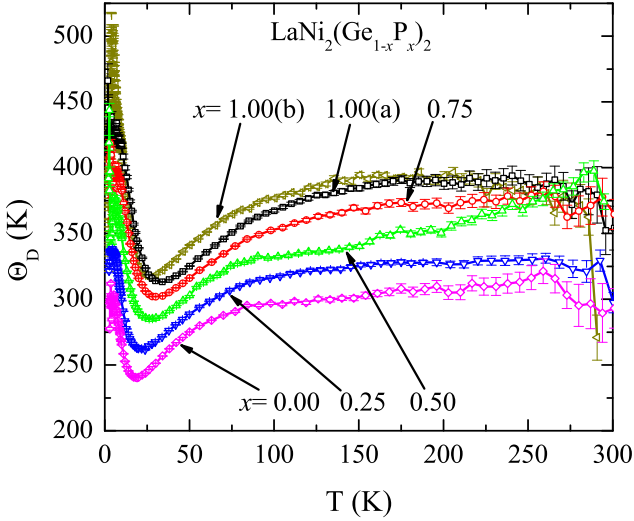}
 	\caption{(Color online) Debye temperature $\Theta_{\rm{D}}$ versus temperature $T$ (open symbols) obtained by solving Eq.~(\ref{eq:Debye}) for each $C_{\rm p}(T)$ data point after subtracting the electronic contribution $\gamma T$. The error bars plotted are calculated using Eq.~(\ref{eq:theta_error}). Solid curves are guides to the eye.}
	\label{fig:Theta_vs_T}
\end{figure}

The value of $\Theta_{\rm{D}}$ was calculated for each data point in Fig.~\ref{fig:Cp_vs_T}, after subtracting the contribution from $\gamma T$ according to Eq.~(\ref{eq:CpSum}), using the Debye function in Eq.~(\ref{eq:Debye}). The resulting $T$ dependences of $\Theta_{\rm D}$ are shown in Fig.~\ref{fig:Theta_vs_T}, where  $\Theta_{\rm{D}}$ is seen to vary nonmonotonically and by up to 30\% with increasing $T$\@. The plots have a similar shape to that for sodium iodide.\cite{Berg1957} The $\Theta_{\rm{D}}$ is expected to be constant below $\Theta_{\rm{D0}}/50$ and above $\Theta_{\rm{D0}}/2$, where $\Theta_{\rm{D0}}$ is the zero-temperature value of $\Theta_{\rm{D}}$.\cite{Gopal1966} Our $\Theta_{\rm D}(T)$ data qualitatively agree with these expectations. 

Considerable scatter in the $\Theta_{\rm D}(T)$ data in Fig.~\ref{fig:Theta_vs_T} occurs at temperatures above 250~K and also below 7~K (although not as clearly visible in the figure).  In these $T$ ranges, $C_{\rm p}$ is becoming nearly independent of $T$, so in these $T$ ranges any error in the value of $C_{\rm V}$ is greatly amplified when $\Theta_{\rm{D}}$ is calculated. The error bars on the values of $\Theta_{\rm D}(T)$ therefore increase significantly in these $T$ regions.  We now consider such errors and for this discussion ignore the difference between $C_{\rm V}$ and $C_{\rm p}$.

The scatter $d\Theta_{\rm D}$ in the derived $\Theta_{\rm D}$ versus $T$ depends on the statistical error $d C_{\rm V}$ in $C_{\rm V}$, 
\be
d\Theta_{\rm{D}}=\frac{dC_{\rm{V}}}{dC_{\rm{V}}/d\Theta_{\rm{D}}}.
\label{eq:theta_error}
\ee
This expression was used to obtain the error bars plotted in Fig.~\ref{fig:Theta_vs_T}. The denominator $dC_{\rm{V}}/d\Theta_{\rm{D}}$ was calculated using the Debye function in Eq.~(\ref{eq:Debye}).

In order to more clearly see why the error in $\Theta_{\rm D}$ substantially increases above $\approx 250$~K, the high-$T$ approximation in Eq.~(\ref{eq:highT_Debye}) can be used, yielding
\begin{displaymath}
C_{\rm{V}} \approx 3R-\frac{3R{\Theta_{\rm{D}}}^2}{20T^2}. \ \ \ \ \ \ \ \ (T\gg \Theta_{\rm{D}})
\end{displaymath}
Taking the derivative with respect to $\Theta_{\rm{D}}$ gives 
\begin{displaymath}
\frac{dC_{\rm{V}}}{d\Theta_{\rm{D}}} \approx \frac{-3R\Theta_{\rm{D}}}{10T^2}. \ \ \ \ \ \ \ \ (T\gg \Theta_{\rm{D}})
\end{displaymath}
Inserting this result into Eq.~(\ref{eq:theta_error}) gives the approximation
\begin{displaymath}
d\Theta_{\rm{D}} \approx -\left(\frac{10 T^2 }{3R\Theta_{\rm{D}}}\right)\,dC_{\rm{V}}. \ \ \ \ \ \ \ \ (T\gg \Theta_{\rm{D}})
\end{displaymath}
This result shows that the error in $\Theta_{\rm{D}}$ is proportional to $T^2$ at high $T$, which results in a dramatic increase in the scatter and error in $\Theta_{\rm D}(T)$ at high $T$.

Using the low-$T$ approximation in Eq.~(\ref{eq:lowT_Debye}) and the same procedure as described above, the error in $\Theta_{\rm{D}}$ is 
\begin{displaymath}
d\Theta_{\rm{D}} \approx -\left(\frac{5 {\Theta_{\rm{D}}}^4}{36  R \pi^4 T^3}\right)\,dC_{\rm{V}}. \ \ \ \ \ \ \ \ (T\ll \Theta_{\rm{D}})
\end{displaymath}
Therefore, similar to the situation at high-$T$, a small error in $C_{\rm p}$ at low~$T$ is greatly amplified when calculating $\Theta_{\rm D}(T)$.

\begin{figure}
	\includegraphics[width=3.3in]{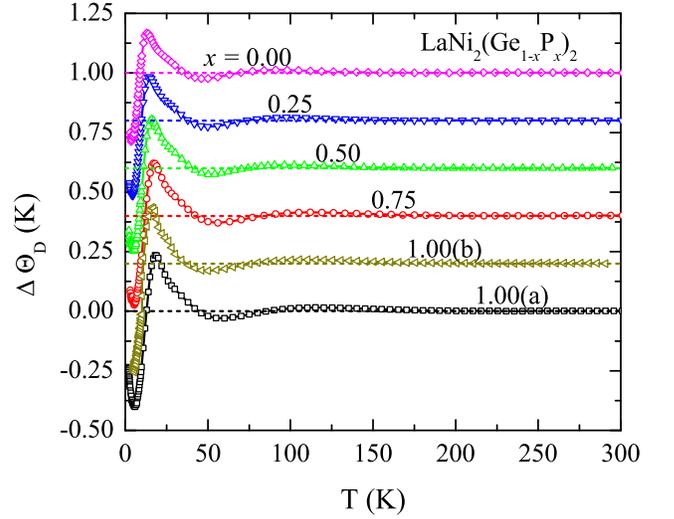}
 	\caption{(Color online) Difference $\Delta\Theta_{\rm D}$ between the $\Theta_{\rm{D}}$ calculated from the Debye function [Eq.~(\ref{eq:Debye})] plotted in Fig.~\ref{fig:Theta_vs_T} and calculated from the Pad\'e approximant [Eqs.~(\ref{eq:Debye_approx}) and (\ref{eq:final_Debye_approx1})]. Solid curves are guides to the eye. For clarity, each plot is offset vertically upwards by 0.2~K from the one below it.  Horizontal dotted lines are at $\Delta \Theta_{\rm D}=0$ for the data set with the corresponding color.}
	\label{fig:Delta_Theta_vs_T}
\end{figure}

To verify the applicability of the Pad\'e approximant for the Debye function developed in Sec.~\ref{Debye_Function}, the $\Theta_{\rm{D}}(T)$ was calculated for each data point in Fig.~\ref{fig:Cp_vs_T} using the Pad\'e approximant in Eq.~(\ref{eq:Debye_approx}) instead of by directly using the Debye function in Eq.~(\ref{eq:Debye}).  The electronic $\gamma T$ contribution was again subtracted from the $C_{\rm p}(T)$ data first. The difference between these $\Theta_{\rm{D}}(T)$ values and those calculated using the Debye function in Eq.~(\ref{eq:Debye_approx}) is plotted versus $T$ in Fig.~\ref{fig:Delta_Theta_vs_T}. The values calculated from the Pad\'e approximant do not deviate by more than 0.35~K from those calculated using the Debye formula.  This error is of order $0.1\%$ of $\Theta_{\rm D}$, which is usually small compared to the error in $\Theta_{\rm D}$ itself and is negligible compared to its $T$ dependence.  Therefore the Pad\'e approximant provides a viable alternative for calculating $\Theta_{\rm{D}}(T)$ that does not require evaluation of the integral in the Debye function~(\ref{eq:Debye}) or the use of a look-up table for each data point.

\subsection{\label{Magnetic} Magnetization and Magnetic Susceptibility Measurements}

\begin{figure}
	\includegraphics[width=3.3in]{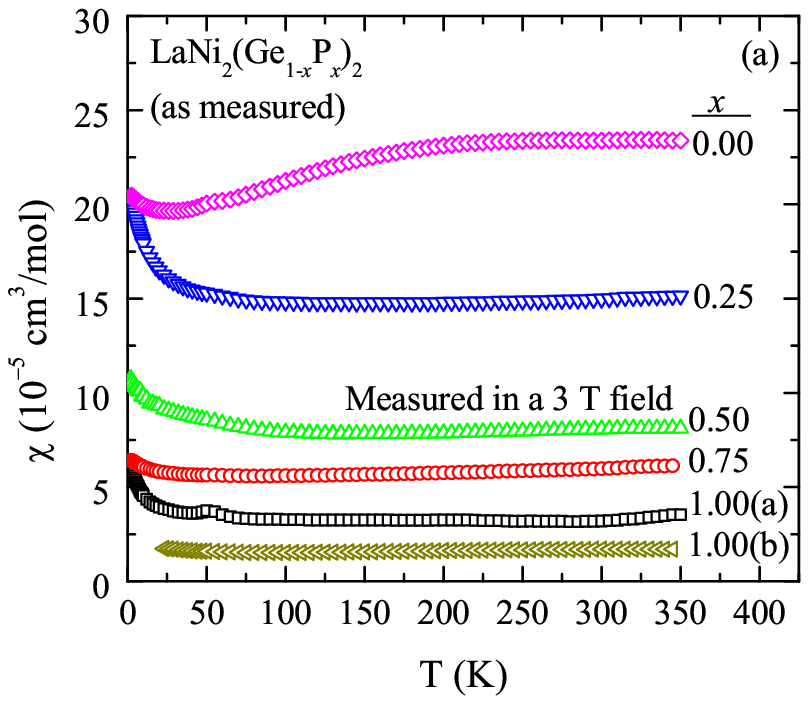}
	\includegraphics[width=3.3in]{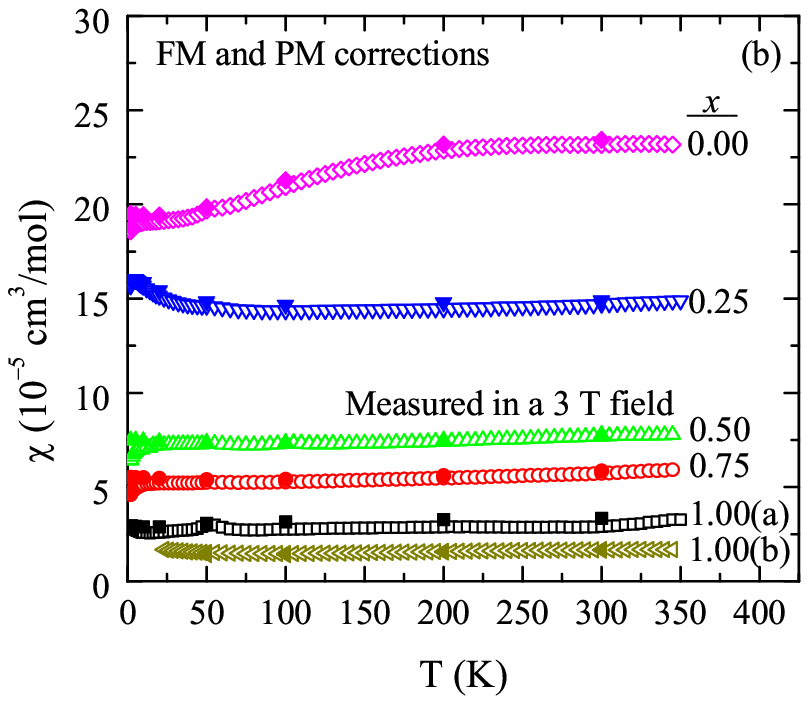}
 	\caption{(Color online) Magnetic susceptibility $\chi$ versus temperature $T$ for the LaNi$_2$(Ge$_{1-x}$P$_x$)$_2$ system. (a) Measured $\chi \equiv M/H$ data (uncorrected). (b) Intrinsic $\chi$ obtained after correcting for both ferromagnetic (FM) and paramagnetic (PM) impurities. Solid symbols of corresponding shape and color are values of $\chi$ found from fitting $M(H)$ isotherms.}
	\label{fig:Chi_vs_T}
\end{figure}

\begin{figure}
	\includegraphics[width=3.3in]{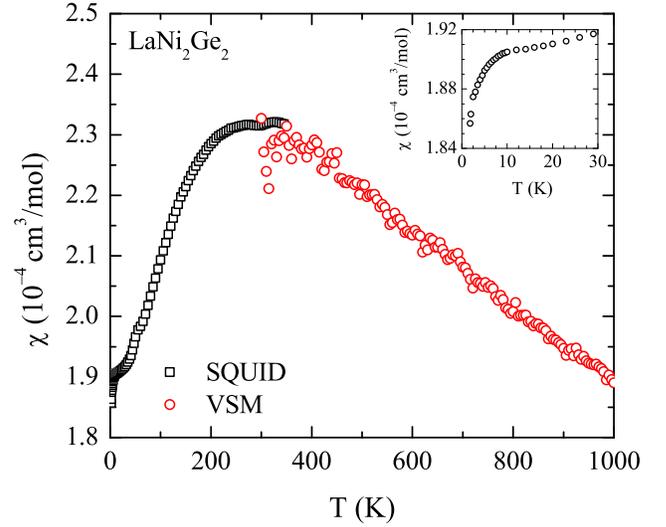}
 	\caption{(Color online) Expanded plot along the vertical axis of the magnetic susceptibility $\chi$ versus temperature $T$ for LaNi$_2$Ge$_2$ up to 1000~K after correction for paramagnetic and ferromagnetic impurity contributions. The data below 350~K were measured with a SQUID magnetometer and the data above 300~K were measured with a VSM, both at applied fields of 3~T\@. Inset: Expanded plot of the SQUID data at low temperatures.  The small downturn below about 10~K is most likely spurious, arising from an imperfect correction for the paramagnetic impurities.}
	\label{fig:LaNi2Ge2_VSM}
\end{figure}

Magnetization versus applied magnetic field $M(H)$ isotherms were measured for the LaNi$_2$(Ge$_{1-x}$P$_x$)$_2$ system for $H=0$--5.5~T and the results are plotted in Figs.~\ref{fig:LaNi2P2_ascast_M_vs_H}--\ref{fig:LaNi2Ge2_M_vs_H} of Appendix~\ref{Sec:MHPlots}. The $M$ versus $T$ for the samples were also measured from 1.8 to 300~K at a fixed field $H = 3$~T and the resulting susceptibilities $\chi \equiv M/H$ are plotted in Fig.~\ref{fig:Chi_vs_T}(a).  As evident from the nonlinear behavior at low fields in the $M(H)$ plots and the upturn that follows the Curie-Weiss-like behavor [$\chi=C/(T-\theta)$] in the observed $\chi(T)$ at low temperatures, we infer the presence of saturating paramagnetic and/or ferromagnetic impurities in the samples. In order to determine the intrinsic behaviors, it is necessary to correct for these impurities. To determine the individual contributions to the susceptibility data, the $M(H)$ curves were fitted by
\be
M(T,H)=M_0+\chi H+f M_{\rm{{sat}}}B_{S} \left[ \frac{g\mu_{\rm{B}} H}{k_{\rm{B}} (T-\theta)} \right],
\label{eq:MH_fit}
\ee
where $M_0$ is the saturation magnetization of the ferromagnetic impurities, $\chi$ is the intrinsic susceptibility of the sample, $f$ is the molar fraction of paramagnetic impurities, $g$ is the spectroscopic splitting factor of the impurities which was fixed at $g=2$ to reduce the number of fitting parameters, $\mu_{\rm{B}}$ is the Bohr magneton, $k_{\rm{B}}$ is Boltzmann's constant, $\theta$ is the Weiss temperature of the paramagnetic impurities (included for consistency with a possible Curie-Weiss law behavior at low $H/T$), $M_{\rm{{sat}}}$ is the saturation magnetization of the paramagnetic impurities, and $B_{S}$ is the Brillouin function. The Brillouin function is
\be
B_{S}(x)=\frac{1}{2S} \left\{ (2S+1)\coth \left[(2S+1)\frac{x}{2} \right] - \coth \left(\frac{x}{2} \right)  \right\},
\label{eq:brillouin}
\ee
where 
\[
x = \frac{g\mu_{\rm{B}} H}{k_{\rm{B}} (T-\theta)}
\]
 and the molar saturation magnetization $M_{\rm{{sat}}}$ of the paramagnetic impurities is
\begin{displaymath}
M_{\rm{{sat}}}=N_{\rm{A}} g S\mu_{\rm{B}},
\end{displaymath}
where $S$ is the spin of the impurities and $N_{\rm{A}}$ is Avogadro's number.

\begin{figure}
\includegraphics[width=3.3in]{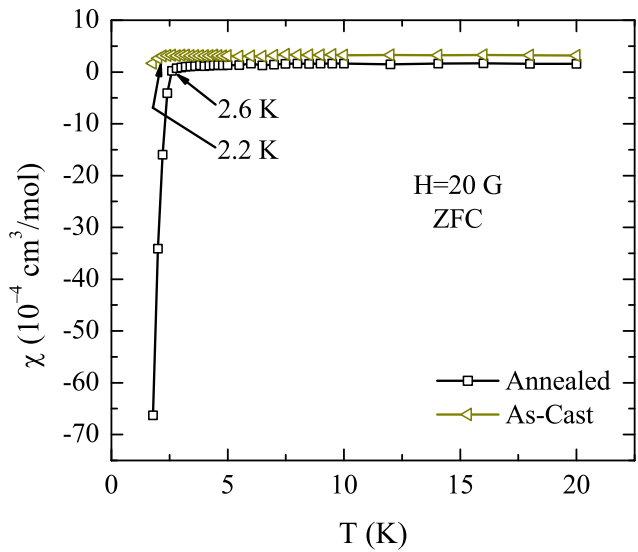}
\caption{(Color online) Zero-field-cooled (ZFC) low-temperature measurement of the magnetic susceptibility $\chi$ versus temperature $T$ for LaNi$_2$P$_2$ with applied field $H=20$~G.}
\label{fig:Chi_SC}
\end{figure}

\begin{table*}
 \caption{\label{table:MH_fitting} Values obtained by simultaneous fitting of the 1.8 and 5~K $M(H)$ isotherms. The parameters listed are $f$ [molar fraction of paramagnetic (PM) impurities], $S$ (spin quantum number of the PM impurities), $\theta$ (Weiss temperature of the PM impurities), $M_0$ (saturation magnetization of the ferromagnetic impurities), and $\chi$ (intrinsic magnetic susceptibility of the compound).  The negative signs of $\theta$ indicate antiferromagnetic interactions between the magnetic impurities.}
 \begin{ruledtabular}	
		\begin{tabular}{l ccccc}
		  & $f$  & $S$ & $\theta$ & $M_0$  & $\chi$ \\
		Sample& $(10^{-5})$  &     &  (K)  &  (G\,cm$^3$/mol) & (10$^{-5}$ cm$^3$/mol)\\ \hline
		LaNi$_2$P$_2$ (annealed)			& 3.89	& 3.48	&$-$5.88	& 0.0478  	& 2.92	\\
		LaNi$_2$(P$_{0.75}$Ge$_{0.25}$)$_2$	& 2.33	& 1.92	&$-$1.02 	& 0.1175	& 5.53	\\
		LaNi$_2$(P$_{0.50}$Ge$_{0.50}$)$_2$ & 4.12	& 2.42	&$-$1.89	& 0.4128	&  	7.49	\\
		LaNi$_2$(P$_{0.25}$Ge$_{0.75}$)$_2$ & 5.73	& 2.57 &$-$1.54	& 0.1834& 16.0	\\
		LaNi$_2$Ge$_2$						& 2.35	& 2.43	&$-$1.08	& 0.0700	& 19.5	\\
		\end{tabular}
 \end{ruledtabular}
\end{table*}

To determine the values of $f$, $S$, and $\theta$ for the paramagnetic impurities, a global two-dimensional surface fit of the $M(H)$ data from 1--5.5~T taken at both 1.8 and 5~K was done using Eq.~(\ref{eq:MH_fit}) with {\tt MATLAB}'s Surface Fitting Tool. For this $H$ range, the ferromagnetic impurities are expected to be nearly saturated, as assumed by Eq.~(\ref{eq:MH_fit}). The parameter values obtained are given in Table~\ref{table:MH_fitting}. Fixing the variables $f$, $S$, and $\theta$ at the respective values for each sample, each $M(H)$ curve at higher temperatures was fitted by Eq.~(\ref{eq:MH_fit}) in the range 1--5.5~T to obtain values for $M_0(T)$ and $\chi(T)$. The $M(H)$ fits are shown in Figs.~\ref{fig:LaNi2P2_ascast_M_vs_H}--\ref{fig:LaNi2Ge2_M_vs_H} in Appendix~\ref{Sec:MHPlots}. A plot of $M_0$ versus $T$ is presented in Fig.~\ref{fig:M0_vs_T} in Appendix~\ref{Sec:MHPlots} and the fitted values of the intrinsic susceptibility $\chi$ are plotted as solid symbols in Fig.~\ref{fig:Chi_vs_T}(b).

In order to correct the $M(T)$ data at $H = 3$~T in Fig.~\ref{fig:Chi_vs_T}(a) for the paramagnetic impurities, the above values of $f$, $S$, and $\theta$ obtained by fitting the $M(H)$ isotherms at 1.8 and 5~K were inserted into the last term of Eq.~(\ref{eq:MH_fit}) to calculate their contributions versus $T$ at $H = 3$~T\@.  These contributions were then subtracted from the respective $M(T)$ data that were already corrected for ferromagnetic impurities to obtain the intrinsic susceptibility of the samples as plotted in Fig.~\ref{fig:Chi_vs_T}(b). Previously reported values of $\chi$ for LaNi$_2$Ge$_2$ are $24\times10^{-5}$~cm$^3$/mol at 300~K (Ref.~\onlinecite{Wernick1982}) and $28\times10^{-5}$~cm$^3$/mol at 296~K (Ref.~\onlinecite{Zell1981}).  The former value is essentially the same as our value $23.2\times10^{-5}$~cm$^3$/mol at 300~K, as seen more clearly in Fig.~\ref{fig:LaNi2Ge2_VSM} below where our $\chi(T)$ data for LaNi$_2$Ge$_2$ are plotted on an expanded vertical scale for temperatures up to 1000~K\@. 

The $\chi(T)$ data in Fig.~\ref{fig:Chi_vs_T} show that the samples in the LaNi$_2$(Ge$_{1-x}$P$_x$)$_2$ system exhibit nearly temperature-independent paramagnetism over the composition region $x=0$.25--1.  This trend does not extend to LaNi$_2$Ge$_2$ ($x=0$), for which a broad maximum appears to occur at $\approx 300$~K, which is close to the upper temperature limit of the SQUID magnetometer. In order to determine whether a maximum in $\chi$ near 300~K does occur, $M(H,T)$ data for the sample of LaNi$_2$Ge$_2$ were measured using a VSM from $T=300$ to~1000~K\@. The $M(H)$ isotherm data are plotted in Fig.~\ref{fig:LaNi2Ge2_M_vs_H}(b) of Appendix~\ref{Sec:MHPlots} and the $\chi(T)\equiv M(T)/H$ data at $H = 3$~T are plotted in Fig.~\ref{fig:LaNi2Ge2_VSM}. The magnetic contributions  due to the sample holder and paramagnetic impurities are corrected for in the plots.

The data in Fig.~\ref{fig:LaNi2Ge2_VSM} clearly show that a broad peak occurs in $\chi(T)$ of LaNi$_2$Ge$_2$ at about 300~K\@.  The plot in Fig.~\ref{fig:Chi_vs_T}(b) shows the peak plotted on a less expanded vertical scale. The cause of this peak is not clear, but may be due to low-dimensional antiferromagnetic correlations.\cite{Johnston1997} Further investigation is needed. The inset in Fig.~\ref{fig:LaNi2Ge2_VSM} shows an expanded view of the low temperature behavior.  Due to its smooth nature, the small downturn in the data below about 10~K is most likely spurious due to a slight error in correcting for the susceptibility contribution of the paramagnetic impurities in the sample.

The onset of superconductivity was observed in both the annealed and as-cast samples of LaNi$_2$P$_2$ at 2.6 and 2.2~K, respectively, from zero-field-cooled (ZFC) magnetization measurements in a field of 20~Oe\@. Figure~\ref{fig:Chi_SC} shows these low-field $\chi(T) \equiv M(T)/H$ measurements. It is not clear whether the data represent  the onset of bulk superconductivity in LaNi$_2$P$_2$ or if the diamagnetism arises from a superconducting impurity phase.  The $\rho(T)$ data in Fig.~\ref{fig:Resistivity} below do not clarify this issue.

We now analyze the normal-state $\chi$ of the samples.  The magnetic susceptibility of a metal consists of the sum of the spin and orbital contributions
\be
\chi= \chi^{\rm spin} + \chi^{\rm orb} .
\label{Eq:chiorbspin}
\ee
In the absence of local magnetic moments, the spin contribution $\chi^{\rm spin}$ is the Pauli susceptibility  $\chi^{\text{Pauli}}$ of the conduction electrons.  One can estimate $\chi^{\text{Pauli}}$ using \cite{Johnston2010}
\be
\chi^{\text{Pauli}}=\frac{g^2}{4} \mu_{\rm{B}}^2 {\cal D}(E_{\rm{F}}), \ \ \ \ \ (T=0)
\label{eq:chi_pauli}
\ee 
where $g$ is the spectroscopic splitting factor and ${\cal D}(E_{\rm{F}})$ is the density of states at the Fermi energy $E_{\rm{F}}$. Setting $g=2$ gives
\be
\chi^{\text{Pauli}}=(3.233 \times 10^{-5}) {\cal D}(E_{\rm{F}}),
\label{eq:chi_pauli_simplified}
\ee
where $\chi^{\text{Pauli}}$ is in units of cm$^3$/mol, ${\cal D}(E_{\rm{F}})$ is in units of states/eV\,f.u.\ for both spin directions and f.u.\ means formula unit.

 \begin{table*}
 \caption{\label{table:magnetic_properties} Contributions to the magnetic susceptibility $\chi$.  The parameters listed are the observed value $\chi^{\text{obs}}$ of $\chi$ averaged over the temperature range measured and the contributions from the Pauli spin susceptibility $\chi^{\rm Pauli}$ and the orbital core susceptibility $\chi^{\rm core}$, Landau susceptibility $\chi^{\rm Landau}$, and Van Vleck susceptibility $\chi^{\rm VV}$. All values are in units of 10$^{-5}$~cm$^3$/mol.}
 \begin{ruledtabular}	
		\begin{tabular}{l D{.}{.}{2} D{.}{.}{5} D{.}{.}{3} D{.}{.}{5} D{.}{.}{5}}
		 Sample & \multicolumn{1}{c}{$\chi^{\text{obs}}$}  & \multicolumn{1}{c}{$\chi^{\text{Pauli}}$} & \multicolumn{1}{c}{$\chi^{\text{core}}$} & \multicolumn{1}{c}{$\chi^{\text{Landau}}$} & \multicolumn{1}{c}{$\chi^{\text{VV}}$} \\ \hline
		LaNi$_2$P$_2$ (as-cast)				& 1.49	& 10.5(3)	& -18.8 & -3.50(1) & 13.3(4)	\\
		LaNi$_2$P$_2$ (annealed)			& 2.82 	& 8.04(3) 	& -18.8 & -2.68(1) & 16.3(4)	\\
		LaNi$_2$(P$_{0.75}$Ge$_{0.25}$)$_2$	& 5.28 	& 10.1(2) 	& -19.3 & -3.37(7) & 17.9(3)	\\
		LaNi$_2$(P$_{0.50}$Ge$_{0.50}$)$_2$ & 7.32 	& 12.7(3) 	& -19.7 & -4.2(1)  & 18.5(4)	\\
		LaNi$_2$(P$_{0.25}$Ge$_{0.75}$)$_2$ & 14.36 & 15.44(8)  & -20.1 & -5.15(3) & 24.7(1)	\\
		LaNi$_2$Ge$_2$ 						& 21.01 & 17.0(3) 	& -20.5 & -5.7(1)  & 30.4(4)
		\end{tabular}
 \end{ruledtabular}
\end{table*}

 \begin{table*}
 \caption{\label{table:resistivity_properties} Values of the Debye temperature determined from resistivity $\rho$ measurements ($\Theta_{\rm{R}}$), the resistivity at the Debye temperature [$\rho(\Theta_{\rm{R}})$] and the residual resistivity ($\rho_0$) obtained from least-squares fits of the $\rho(T)$ data in Fig.~\ref{fig:Resistivity}(a) by Eq. (\ref{eq:Gruneisen_approx_final}). Also listed are $\rho$ values at $\sim 2$~K and at 300~K, and literature values parallel to the $c$-axis ($\rho_c$) and parallel to the $a$-axis ($\rho_a$) of a single crystal.  The systematic errors in our $\rho$ values due to uncertainties in the geometric factors are of order 10\%.}
 \begin{ruledtabular}	
		\begin{tabular}{l l D{.}{.}{5} D{.}{.}{5} c c l}
		 Sample 	& $\Theta_{\rm{R}}$ (K) & \multicolumn{1}{c}{$\rho(\Theta_{\rm{R}})$} & \multicolumn{1}{c}{$\rho_0$}			& $\rho$($\sim 2$ K)	& $\rho$(300 K)		& Ref.	\\
					& 						& \multicolumn{1}{c}{($\mu \Omega$\,cm)} & \multicolumn{1}{c}{($\mu \Omega$\,cm)}  	& ($\mu \Omega$\,cm)	& ($\mu\Omega$\,cm)	&		\\ \hline
		LaNi$_2$P$_2$  (as-cast)			& 211(2)	& 44.9(3)	& 83.15(3)	& 83		& 148		& This work	\\
		\ \ \ \ \ \ \ \ \ \ (annealed)		& 265(3) 	& 109.(1)	& 26.0(1)	& 25		& 152		& This work	\\
		LaNi$_2$(P$_{0.75}$Ge$_{0.25}$)$_2$	& 242(1)	& 85.8(4)	& 191.51(5) & 191						& 300		& This work	\\
		LaNi$_2$(P$_{0.50}$Ge$_{0.50}$)$_2$ & 208(1)	& 102.5(7)	& 249.79(9) & 249						& 401		& This work	\\
		LaNi$_2$(P$_{0.25}$Ge$_{0.75}$)$_2$ & 119(7)	& 37.(2)	& 95.9(3) 	& 95						& 189		& This work	\\
		LaNi$_2$Ge$_2$	 					& 148(5)	& 38.(1)	& 6.8(2)   	& 6.1						& 85		& This work	\\
											&			&			&				& 0.4						&				& \onlinecite{Maezawa1999}\\
											&			&			&				& $\sim 1$						& $\sim 26$ ($\rho_a$), $\sim 38$ ($\rho_c$)	& \onlinecite{Fukuhara1995} \\
											&			&			&				& $\sim 5$						& $\sim 80$			& \onlinecite{Knopp1988} \\
											& 			&			&				& $\sim 1$--2						& $\sim 43$			& \onlinecite{Schneider1983}
		\end{tabular}	
 \end{ruledtabular}
\end{table*}
To calculate $\chi^{\rm Pauli}$, one can obtain an estimate of ${\cal D}(E_{\rm{F}})$ from the Sommerfeld electronic linear specific heat coefficient $\gamma$ according to
\begin{displaymath}
\gamma=\gamma_0 (1+\lambda_{\rm{ep}}),
\end{displaymath}
where\cite{Kittel2005}
\bea
\gamma_0 = \frac{\pi^2k_{\rm B}^2}{3}{\cal D}(E_{\rm{F}}) = 2.359\, {\cal D}(E_{\rm{F}})
\label{Eq:DfromGamma0}
\eea
is the bare Sommerfeld coefficient in the absence of electron-phonon coupling, $\lambda_{\rm{ep}}$ is the electron-phonon coupling constant, and in the right-hand equality of Eq.~(\ref{Eq:DfromGamma0}) $D(E_{\rm{F}})$ is in units of states/eV\,f.u.\ for both spin directions and $\gamma_0$ is in units of mJ/mol\,K$^2$. Taking $\lambda_{\rm{ep}} = 0$ yields
\be
\chi^{\text{Pauli}}=(1.370\times10^{-5})\gamma,
\label{eq:chi_pauli_simplified2}
\ee
where $\chi^{\text{Pauli}}$ is in units of cm$^3$/mol and $\gamma$ is in units of mJ/mol\,K$^2$. The values of ${\cal D}(E_{\rm{F}})$ and $\chi^{\text{Pauli}}$ obtained using Eqs.~(\ref{Eq:DfromGamma0}) and (\ref{eq:chi_pauli_simplified2}), respectively, and using the $\gamma$ values given in Table~\ref{table:HC_properties}, are listed in Table~\ref{table:magnetic_properties}.

The orbital susceptibility consists of three contributions
\be
\chi^{\text{orb}}=\chi^{\rm core}+\chi^{\text{VV}}+\chi^{\text{Landau}},
\label{Eq:chiorb}
\ee
where $\chi^{\text{core}}$ is the diamagnetic contribution from the atomic core electrons, $\chi^{\text{VV}}$ is the paramagnetic Van Vleck susceptibility, and $\chi^{\text{Landau}}$ is the diamagnetic Landau susceptibility of the conduction electrons. The Landau susceptibility is approximated by\cite{Elliot1998}
\begin{displaymath}
\chi^{\text{Landau}}=-\frac{1}{3} \left(\frac{m_{\rm e}}{m^*} \right)^2 \chi^{\text{Pauli}}.
\end{displaymath}
For our samples, it is assumed that $m^*=m_{\rm e}$, therefore $\chi^{\text{Landau}}$ can be calculated from the $\chi^{\text{Pauli}}$ obtained above. An estimate of $\chi^{\rm core}$ was obtained from the sum of the Hartree-Fock diamagnetic atomic susceptibilities.\cite{Mendelsohn1970}  Using the observed average value of the susceptibility ($\chi^{\text{obs}}$) over the temperature range 2--300~K and the calculated sum of the values $\chi^{\text{Pauli}}$, $\chi^{\text{core}}$ and $\chi^{\text{Landau}}$, the Van Vleck susceptibility is calculated from Eqs.~(\ref{Eq:chiorbspin}) and~(\ref{Eq:chiorb}) via
\begin{displaymath}
\chi^{\text{VV}}=\chi^{\text{obs}}-(\chi^{\text{Pauli}}+\chi^{\text{core}}+\chi^{\text{Landau}}).
\end{displaymath}
The values obtained in the manner described for $\chi^{\text{Pauli}}$, $\chi^{\text{core}}$, $\chi^{\text{VV}}$ and $\chi^{\text{Landau}}$ are listed in Table~\ref{table:magnetic_properties}.

\subsection{\label{Transport} Electrical Resistivity Measurements}

\begin{figure}
	\includegraphics[width=3.3in]{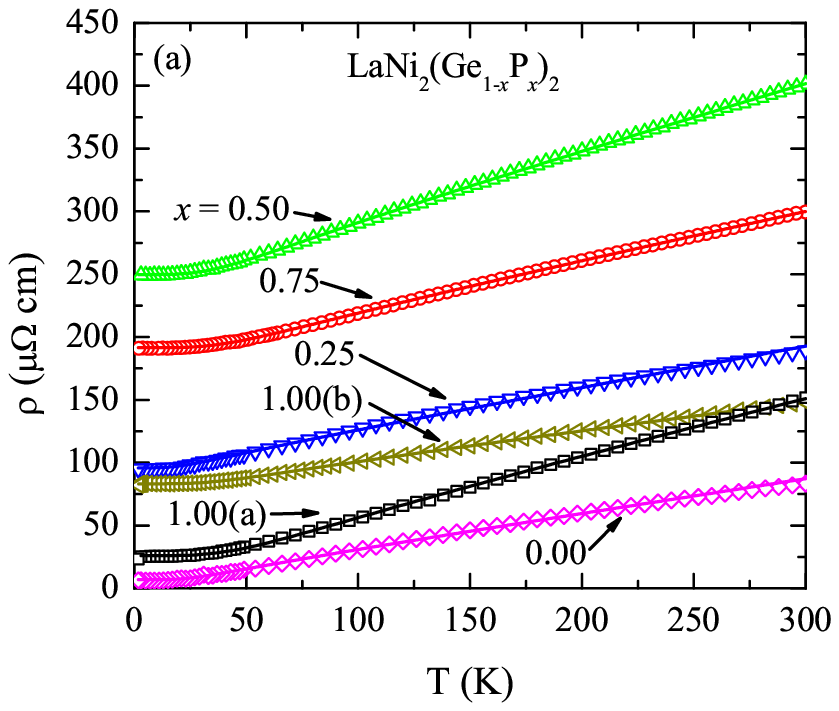}
	\includegraphics[width=3.2in]{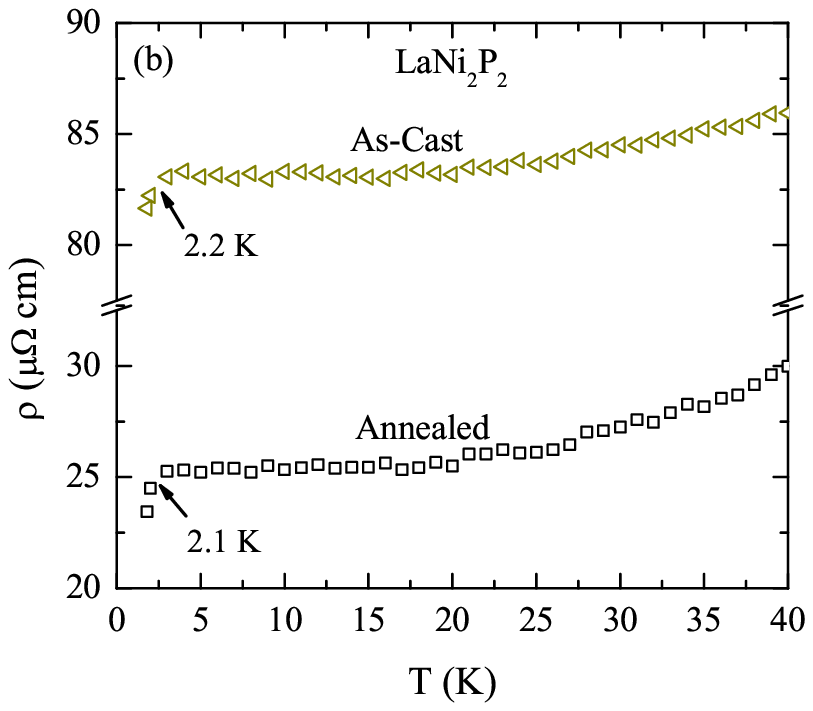}
 	\caption{(Color online) (a) Electrical resistivity $\rho$ versus temperature $T$ for the LaNi$_2$(Ge$_{1-x}$P$_x$)$_2$ system (open symbols). Solid curves of corresponding color are least-squares fits by Eq.~(\ref{eq:Gruneisen_approx_final}), which includes the Pad\'e approximant in Eq.~(\ref{eq:Gruneisen_approx}) that is used in place of the normalized Bloch-Gr\"uneisen function in Eq.~(\ref{eq:norm_Gruneisen2}).  Values of the fitting parameters obtained are listed in Table~\ref{table:resistivity_properties}. (b) Expanded plots at low temperatures of $\rho(T)$ for as-cast and annealed LaNi$_2$P$_2$.}
	\label{fig:Resistivity}
\end{figure}

The $\rho$ versus $T$ data for our LaNi$_2$(Ge$_{1-x}$P$_{x}$)$_2$ samples are plotted in Fig.~\ref{fig:Resistivity}(a).  The magnitudes of the data may not be reliable due to the polycrystalline nature of the samples and the resulting grain boundary scattering. Such scattering is expected and found to be smallest for the samples of LaNi$_2$Ge$_2$ and LaNi$_2$P$_2$ that were cut from arc-melted buttons and therefore had a higher density than the other samples cut from sintered pellets. The values of $\rho$ at 1.8 and 300~K are listed in Table~\ref{table:resistivity_properties} along with previously reported values from the literature.\cite{Maezawa1999, Fukuhara1995, Knopp1988, Schneider1983}  The positive slopes of the $\rho(T)$ data indicate that all samples are metallic.  The maximum resistivities at 1.8~K and at 300~K both occur for $x=0.50$ where the disorder on the Ge/P sublattice is largest, as might have been anticipated, so this may be an intrinsic effect.

As seen in the expanded plot at low temperatures down to our low-$T$ limit of 1.8~K in Fig.~\ref{fig:Resistivity}(b), there appears to be an onset of superconductivity occurring at $\approx 2.1$~K for both the annealed and as-cast samples of LaNi$_2$P$_2$, consistent with the above measurements of the low-field magnetic shielding susceptibilities in Fig.~\ref{fig:Chi_SC}. It is not known if these results indicate the onset of bulk superconductivity or whether they are due to a superconducting impurity phase.

The $T$ dependence of $\rho$ was fitted by Eq.~(\ref{eq:Gruneisen_approx_final}), which includes our Pad\'e approximant for the Bloch-Gr\"uneisen function.  From these fits shown in Fig.~\ref{fig:Resistivity}(a), estimates of the Debye temperature ($\Theta_{\rm{R}}$), $\rho$ at the Debye temperature [$\rho(\Theta_{\rm{R}})$], and residual resistivity ($\rho_0$) were obtained as  listed in Table~\ref{table:resistivity_properties}.  The $\Theta_{\rm R}$ for the ${\rm LaNi_2(P_{0.25}Ge_{0.75})_2}$ sample is lower than the values for the adjacent compositions.  We speculate that this arises from inaccuracies introduced by the polycrystalline nature of the samples.

The $\Theta_{\rm R}$ obtained from $\rho(T)$ measurements is usually different from that obtained by heat capacity measurements ($\Theta_{\rm D}$) although in some cases agreement found.\cite{Gopal1966}  For our samples in the LaNi$_2$(Ge$_{1-x}$P$_x$)$_2$ system, while the magnitudes of $\Theta_{\rm R}$ differ considerably from sample to sample, both sets of $\Theta_{\rm R}$ and $\Theta_{\rm D}$ data in Tables~\ref{table:resistivity_properties} and~\ref{table:HC_properties}, respectively, indicate an overall decrease in the Debye temperature with increasing Ge content. 

Like the Debye model, the BG model makes several assumptions and approximations in order to obtain an analytic formula for $\rho(T)$:
\begin{itemize}
\item{A strong approximation is that the lattice vibrations have a Debye spectrum.}
\item{Another strong approximation is that Umklapp conduction electron scattering processes are ignored.  Umklapp scattering is expected to be $T$-dependent in our $T$ range.\cite{Ziman1960}}
\item{Only longitudinal phonons are assumed to contribute to electron-phonon scattering.}
\item{The calculation is carried out at constant volume instead of at constant pressure as in experiments.}
\item{The conduction electron energy is assumed not to change due to scattering off phonons.} \item{The phonons are assumed to be in thermal equilibrium.}
\end{itemize}
As a result of these assumptions and approximations, the $T$ dependence of $\Theta_{\rm{R}}$ is much stronger than that of $\Theta_{\rm{D}}$.\cite{Ziman1960}  Zimon concluded, ``The actual observed value of $\Theta_{\rm R}$ does not have any great significance.''\cite{Ziman1960}  Thus $\Theta_{\rm{R}}$ is less meaningful than $\Theta_{\rm{D}}$.  Furthermore, differences between $\Theta_{\rm{R}}$ and $\Theta_{\rm{D}}$ are not unexpected, given the different assumptions of the BG and Debye models enumerated here and in Sec.~\ref{HC}, respectively.

Now we compare our experimental result for the magnitude of $\rho$(300~K) for the annealed sample of ${\rm LaNi_2Ge_2}$ with the value predicted by the Bloch-Gr\"uneisen theory.  We chose to use the $\rho$ data for this compound for comparison with the theory because it has the lowest residual resistivity of our five samples.  The BG theory describes monatomic materials. In order to use this theory to predict the resistivity of polyatomic compounds such as ${\rm LaNi_2Ge_2}$, it is necessary to slightly modify Eq.~(\ref{eq:R}) as
\begin{equation}
 \mathcal{R}(\Theta_{\rm R}) =\frac{\hbar}{e^2} \left[ \frac{\pi^3 (3 \pi^2)^{1/3} \hbar^2}{4 n_{\rm{cell}}^{2/3} a k_{\rm{B}} \Theta_{\rm{R}}} \left(\frac{1}{M}\right)_{\rm ave} \right],
 \label{eq:R_mod}
\end{equation}
where the variables have the same meaning as before, except
\begin{equation}
\left( \frac{1}{M} \right)_{\rm ave} = \frac{N_{\rm A}}{n}\sum\limits_{i=1}^n \frac{1}{M_i}
\end{equation}
is the average inverse mass of the atoms in a f.u., $n$ is the number of atoms per f.u.\ and $M_i$ is the atomic weight of element~$i$.  In this case, $a=(V_{\rm cell}/nZ)^{1/3}$ because there are $Z=2$ f.u.\ of ${\rm LaNi_2Ge_2}$ with $n=5$ per body-centered tetragonal unit cell with volume $V_{\rm cell} = a^2c$ where $a$ and $c$ are given in Table~\ref{table:structure}.

The number of carriers per f.u.\@ in LaNi$_2$Ge$_2$ predicted by band structure calculations\cite{Yamagami1999} is 1.54. Therefore, $n_{\rm cell}=1.54/5 = 0.308$~carriers/atom because the carriers are modeled as being evenly distributed among the 5 atoms in the f.u. As described previously, to calculate ${\cal R}(\Theta_{\rm R})$ in units of $\Omega$\,cm in Eq.~(\ref{eq:R_mod}), the prefactor $\hbar/e^2$ is set to 4108.24 $\Omega$ and the part in the square brackets is calculated in cgs units. Using $\Theta_{\rm R}=148$~K from Table~\ref{table:resistivity_properties}, $\mathcal{R}(\Theta_{\rm R})$ is calculated to be 3.763~$\mu\Omega$\,cm. Inserting this result into Eq.~(\ref{eq:Gruneisen}) for $T=300$~K, $\rho(300$~K) is predicted to be 7.53 $\mu\Omega$\,cm.  This value is about a factor of 10 smaller than the experimental value of 79~$\mu\Omega$\,cm [after subtracting $\rho(1.8$~K)] from Table~\ref{table:resistivity_properties}, which is a large disagreement in magnitude even though the $T$ dependence for $\rho$ for ${\rm LaNi_2Ge_2}$ was well-fitted by the BG theory.  This large disagreement is typical of the BG theory as applied to transition metals and alkaline-earth metals\cite{Blatt1968} and illustrates why it is necessary to use an adjustable prefactor $\rho(\Theta_{\rm R})$ in Eq.~(\ref{eq:Gruneisen_approx_final}) in addition to the fitting parameter $\Theta_{\rm R}$ in order to fit both the magnitude and $T$ dependence of experimental data using the Bloch-Gr\"uneisen model.

The reason for the enhanced electron-lattice resistivity in transition metals with both $s$- and $d$-bands crossing the Fermi energy is that the $s$-electrons carry most of the current due to their much lower effective mass than the $d$-electrons, and electron-phonon scattering from an $s$-band into a much higher density of states $d$-band acts as a conduction electron sink, thus strongly enhancing the resistivity due to electron-phonon scattering.\cite{Ziman1960}

\section{\label{Conclusion} Summary and Conclusions}

Single- or nearly single-phase polycrystalline samples of LaNi$_2$(Ge$_{1-x}$P$_x$)$_2$ with the compositions $x=0$, 0.25, 0.50, 0.75, and 1 were synthesized and their properties measured.  Rietveld refinements of the powder XRD patterns showed that all samples have the tetragonal ThCr$_2$Si$_2$ structure with space group \emph{ I}4/\emph{mmm}. The refined crystal data are presented in Table~\ref{table:structure}. A possible stacking disorder was observed along the $c$-axis in the as-cast sample of LaNi$_2$P$_2$ as revealed by the $c$-axis line broadening in Fig.~\ref{fig:LaNi2P2_comparison_XRD}.

Electrical resistivity $\rho$ measurements showed a positive temperature coefficient for all samples from 2~K to 300~K, indicating that all compositions in this system are metallic.  Consistent with this result, the low-$T$ $C_{\rm p}$ measurements yield a rather large Sommerfeld electronic specific heat coefficient $\gamma = 12.4(2)$~mJ/mol\,K$^2$ for $x = 0$ reflecting a large density of states at the Fermi energy comparable to the largest values found for the $A$Fe$_2$As$_2$ class of materials with the same crystal structure.\cite{Johnston2010}  The $\gamma$ decreases approximately linearly with $x$ to 7.4(1)~mJ/mol\,K$^2$ for $x=1$.  

New Pad\'e approximants for the Debye and Bloch-Gr\"uneisen functions were presented. They have the distinct advantage of being able to precisely reproduce the power law dependences at both the low- and high-$T$ limits of the function they are approximating. The Pad\'e approximants presented here for each of $C_{\rm V}(T)$ and $\rho(T)$ cover the entire $T$ range and have a good balance of high accuracy and low number of terms.  The $T$ dependences of $\rho$ for all samples were well-fitted by the Bloch-Gr\"uneisen model and values of the Debye temperature $\Theta_{\rm R}$ were obtained (Table~\ref{table:resistivity_properties}), although the measured magnitudes were larger than calculated on the basis of this model.  Fitting the $T$ dependences of $C_{\rm p}$ revealed significant $T$ dependences of the Debye temperatures $\Theta_{\rm D}$.

Magnetization and magnetic susceptibility measurements of our LaNi$_2$(Ge$_{1-x}$P$_x$)$_2$ samples revealed nearly temperature-independent paramagnetic behavior (except for LaNi$_2$Ge$_2$), with increasing susceptibility as the Ge concentration is increased.  For  LaNi$_2$Ge$_2$, a broad peak was observed in $\chi(T)$ at $\approx 300$~K\@. A possible explanation is that the peak arises from the onset of strong antiferromagnetic correlations in a quasi-two-dimensional magnetic system.\cite{Johnston1997}  There may be a correlation here between the $\chi(T)$ behavior and the corrugated electron cylinder Fermi surface found by Yamagami\cite{Yamagami1999} that was mentioned in the Introduction.  However, the heat capacity data were well-fitted from 1.8 to 300~K for all samples by an electronic $\gamma T$ term plus a lattice contribution described by the Debye model as shown in Fig.~\ref{fig:Cp_vs_T} and Table~\ref{table:HC_properties}, with no clear evidence for an additional magnetic contribution in LaNi$_2$Ge$_2$. 

An interesting finding is the onset of a superconducting transition at $T \approx 2.2$~K in both the annealed and as-cast samples of LaNi$_2$P$_2$. This onset was observed in both the resistivity and magnetic susceptibility measurements. However, it is not clear if these onsets are due to bulk superconductivity or to an impurity phase. Apart from the broad peak in $\chi(T)$ for LaNi$_2$Ge$_2$ and the potential bulk superconductivity in LaNi$_2$P$_2$, there were no other signs of structural, magnetic, or superconducting transitions down to 1.8~K in the samples.

\acknowledgments

Work at the Ames Laboratory was supported by the Department of Energy-Basic Energy Sciences under Contract No.~DE-AC02-07CH11358.

%\clearpage
\appendix

\section{\label{Pade} Pad\'e Approximants}

\subsection{\label{AppBG} Bloch-Gr\"uneisen Model}

In order to construct a Pad\'e approximant function that accurately represents another function, the power law $T$ dependences of the latter function must be computed at high and low temperatures and the coefficients of the Pad\'e approximant adjusted so that both of these limiting $T$ dependences are exactly reproduced (to numerical precision). For the BG function at high $T$, $\Theta_R/T \ll 1$ in Eq.~(\ref{eq:norm_Gruneisen2}). Therefore, the integrand in Eq.~(\ref{eq:norm_Gruneisen2}) can be expanded in a Taylor series about $x=0$ as
\begin{displaymath}
\frac{x^5}{(e^x-1)(1-e^{-x})}\approx x^3-\frac{x^5}{12}+\mathcal{O}(x^7). \ \ \ \ (x\ll 1)
\end{displaymath}
Equation~(\ref{eq:norm_Gruneisen2}) then becomes
\bea
\rho_{\rm{n}}(T_{\rm{n}})  &\approx&  A T_{\rm{n}}^1+ 0\,T_{\rm{n}}^0 + B {T_{\rm{n}}}^{-1}+{\cal O}(T_{\rm{n}}^{-3}) \ \ (T\gg\Theta_{\rm{R}})\nonumber\\*
A&=&1.056\,565 \label{eq:highT_Gruneisen} \\*
B&=&-0.058\,698\,04 \nonumber
\eea
At low temperatures, $\Theta_R/T \rightarrow \infty$ and the upper limit to the integral in Eq.~(\ref{eq:norm_Gruneisen2}) can be set to~$\infty$.  The integral in Eq.~(\ref{eq:norm_Gruneisen2}) is then
\begin{displaymath}
\int_0^\infty{\frac{x^5}{(e^x-1)(1-e^{-x})}dx} = 5!\zeta(5) \approx 124.431\,331,
\end{displaymath}
where $\zeta(z)$ is the Riemann zeta function.  Inserting this result into Eq.~(\ref{eq:norm_Gruneisen2}) yields
\bea
\rho_{\rm{n}} &=& C {T_{\rm{n}}}^5.\ \ \ \ \ \ \ \ (T\ll\Theta_{\rm{R}})
\label{eq:lowT_Gruneisen} \\*
C &=& 525.8790 \nonumber
\eea
There are no additional terms in powers of $T_{\rm{n}}$ in the Taylor series expansion of $\rho_{\rm{n}}(T_{\rm{n}})$ about $T_{\rm{n}}=0$, because it is not possible to express the exponentials in the integral in Eq.~(\ref{eq:norm_Gruneisen2}) as Taylor series in $1/x$ about $1/x=0$.
	
Based on the above low- and high-$T$ expansions of the normalized BG function, the coefficients of the Pad\'e approximant are now chosen so that the limiting $T$ dependences of the approximant exactly match the required power law $T$ dependences in Eqs.~(\ref{eq:highT_Gruneisen}) and (\ref{eq:lowT_Gruneisen}). In addition, intermediate power terms were added in pairs (one in the numerator and one in the denominator) until there were enough for the approximant to accurately fit the intermediate temperature range of the BG function. The resulting approximant has the form
\[
\rho_{\rm{n}}(T_{\rm n}) = \frac{N_0+\frac{N_1}{T_n}+\frac{N_2}{{T_n}^2}+\frac{N_3}{{T_n}^3}}{\frac{D_1}{T_n}+\frac{D_2}{{T_n}^2}+\cdots +\frac{D_7}{{T_n}^7}+ \frac{D_8}{{T_n}^8}}.\hspace{0.9in}(\ref{eq:Gruneisen_approx})
\]

In order that Taylor series expansions of the Pad\'e approximant~(\ref{eq:Gruneisen_approx}) at high- and low-$T$ correctly reproduce the coefficients of the limiting $T$ dependences of the normalized BG function in Eqs.~(\ref{eq:highT_Gruneisen}) and~(\ref{eq:lowT_Gruneisen}), some of the coefficients $N_i$ and $D_i$ in Eq.~(\ref{eq:Gruneisen_approx}) are not independent.  Expanding Eq.~(\ref{eq:Gruneisen_approx}) as a Taylor series in $1/T_{\rm n}$ about $1/T_{\rm n} = 0$ gives
\bea
\rho_{\rm n}(T_{\rm n}) &=& \frac{N_0}{D_1} T_{\rm n}^1 + \frac{-D_2 N_0 + D_1 N_1}{{D_1}^2}\,T_{\rm n}^0\nonumber \\
&+&  \frac{{D_2}^2 N_0-D_1 D_3 N_0 - D_1 D_2 N_1 + {D_1}^2 N_2}{{D_1}^3} {T_{\rm n}}^{-1}\nonumber \\
&+&  \mathcal{O}({T_{\rm n}}^{-2}). 
\label{eq:Gruneisen_Pade_High}
\eea
Equating the coefficients in Eq.~(\ref{eq:Gruneisen_Pade_High}) with the respective coefficients in the high-$T$ Taylor series expansion of the normalized BG function in Eq.~(\ref{eq:highT_Gruneisen}) yields
\be
D_1=\frac{N_0}{A},\ D_2=\frac{N_1}{A},\ D_3=\frac{A N_2 - B N_0}{A^2}.
\label{eq:Gruneisen_highT_coeff}
\ee
The Taylor series expansion of the approximant~(\ref{eq:Gruneisen_approx}) about $T_{\rm n}=0$ is
\be
\rho_{\rm n} = \frac{N_3}{D_8} {T_{\rm n}}^5 + \mathcal{O}({T_{\rm n}}^6). \nonumber
\ee
Equating the coefficient of this $T^5$ term with that of the Bloch-Gr\"uneisen function in Eq.~(\ref{eq:lowT_Gruneisen}) yields 
\be
D_8=\frac{N_3}{C}.
\label{eq:Gruneisen_lowT_coeff}
\ee
Using the four constraints in Eqs.~(\ref{eq:Gruneisen_highT_coeff}) and~(\ref{eq:Gruneisen_lowT_coeff}), one can exactly reproduce the ${T_{\rm n}}^1$, ${T_{\rm n}}^0$, and ${T_{\rm n}}^{-1}$ dependences of the high-$T$ expansion in Eq.~(\ref{eq:highT_Gruneisen}) and the ${T_{\rm n}}^5$ dependence of the low-$T$ expansion in Eq.~(\ref{eq:lowT_Gruneisen}) for the normalized BG function.

A table of values generated from Eq.~(\ref{eq:norm_Gruneisen2}) and plotted above in Fig.~\ref{fig:Gruneisen_Fit}(a) was least-squares fitted by Eq.~(\ref{eq:Gruneisen_approx}) using the constraints in Eqs.~(\ref{eq:Gruneisen_highT_coeff}) and~(\ref{eq:Gruneisen_lowT_coeff}) for $D_1$, $D_2$, $D_3$, and $D_8$.  The final values of the coefficients in the Pad\'e approximant (\ref{eq:Gruneisen_approx}) are listed in Table~\ref{table:Gruneisen_approx}.  

\subsection{\label{PadeDebye} Debye Model}

At high temperatures $1/T_{\rm{n}}\ll 1$, the integrand in Eq.~(\ref{eq:norm_Debye}) can be expanded in a Taylor series about $x=0$, yielding
\begin{displaymath}
\frac{x^4 e^x}{(e^x-1)^2} = x^2-\frac{x^4}{12}+\mathcal{O}(x^6).
\end{displaymath}
After evaluating the integral in Eq.~(\ref{eq:norm_Debye}) using this approximation, the $ C_{\rm{n}}(T_{\rm{n}})$ becomes
\be
C_{\rm{n}}(T_{\rm{n}}) = ET^0+0\,T^{-1}+F {T_{\rm{n}}}^{-2}+\mathcal{O}({T_{\rm{n}}}^{-4}) \ \  (T\gg\Theta_{\rm{D}})
\label{eq:highT_Debye}
\ee
where
\bea
E&=&3,\nonumber \\
F&=&-\frac{3}{20}.\nonumber
\eea
In the limit of low temperatures $1/T_{\rm{n}} \to \infty$, the integral in Eq.~(\ref{eq:norm_Debye}) can be evaluated  with an upper limit of $\infty$ to obtain
\bea
C_{\rm{n}}(T_{\rm{n}})&=&G {T_{\rm{n}}}^3 \ \ \ \ \ \ \ \ (T\ll\Theta_{\rm{D}})\nonumber \\
G&=&\frac{12 \pi^4}{5}.
\label{eq:lowT_Debye}
\eea
This $T^3$ dependence of the lattice heat capacity at low temperatures is universal and is known as the Debye $T^3$ law. Similar to the Bloch-Gr\"uneisen function, it is not possible to obtain additional terms in the Taylor series expansion of $C_{\rm{n}}(T_{\rm{n}})$ in Eq.~(\ref{eq:norm_Debye}) about $T_{\rm{n}} = 0$. 

Since the Pad\'e approximant must follow Eqs.~(\ref{eq:highT_Debye}) and (\ref{eq:lowT_Debye}) at high and low temperatures, respectively, the approximant was set up in the form
\[
C_{\rm{n}}(T_{\rm{n}}) = \frac{N_0+\frac{N_1}{{T_{\rm{n}}}}+\frac{N_2}{{T_{\rm{n}}}^2}+\cdots+\frac{N_5}{{T_{\rm{n}}}^5}}{D_0+\frac{D_1}{{T_{\rm{n}}}}+\frac{D_2}{{T_{\rm{n}}}^2}+\cdots +\frac{D_7}{{T_{\rm{n}}}^7}+\frac{D_8}{{T_{\rm{n}}}^8}}.\hspace{0.45in}(\ref{eq:Debye_approx})
\]
The number of intermediate power terms was increased in pairs (one in the numerator and one in the denominator) until the final fitted approximant accurately matched the Debye function in the intermediate $T$ range.  Using the same procedure as described in Sec.~\ref{Gruneisen}, the high- and low-$T$ limits of the Debye function in Eqs.~(\ref{eq:highT_Debye}) and (\ref{eq:lowT_Debye}) and the corresponding Taylor series expansions of the Pad\'e approximant~(\ref{eq:Debye_approx}) yield the constraints
\bea
D_0&=&\frac{N_0}{E} \label{Eq:DebyeNDConstraints} \\*
D_1&=&\frac{N_1}{E} \nonumber \\*
D_2&=&\frac{-F N_0 + E N_2}{E^2} \nonumber \\*
D_8&=&\frac{N_5}{G}. \nonumber
\eea
These constraints on $D_0$, $D_1$, $D_2$ and $D_8$ guarantee that the ${T_{\rm n}}^0$, ${T_{\rm n}}^{-1}$, and ${T_{\rm n}}^{-2}$ terms in the high temperature series expansion and the ${T_{\rm n}}^3$ term in the low-$T$ expansion of the Pad\'e approximant~(\ref{eq:Debye_approx}) exactly match to within numerical accuracy the corresponding terms in the Taylor series expansion of the Debye function in Eqs.~(\ref{eq:highT_Debye}) and~(\ref{eq:lowT_Debye}), respectively.

The Debye function data in Fig.~\ref{fig:Debye_Fit}(a) were least-squares fitted by the Pad\'e approximant in Eq.~(\ref{eq:Debye_approx}) using the constraints in Eqs.~(\ref{Eq:DebyeNDConstraints}). After fitting, the denominator of the approximant was checked for zeros and all were found to be at non-real $T_{\rm{n}}$. This insures that the approximant does not diverge at any (real positive) temperature.  The final coefficients in the Pad\'e approximant are listed in Table~\ref{table:Debye_coeff}.

\section{\label{RietveldPlots} Rietveld Refinement Figures}

\begin{figure}
	\includegraphics[width=3.3in]{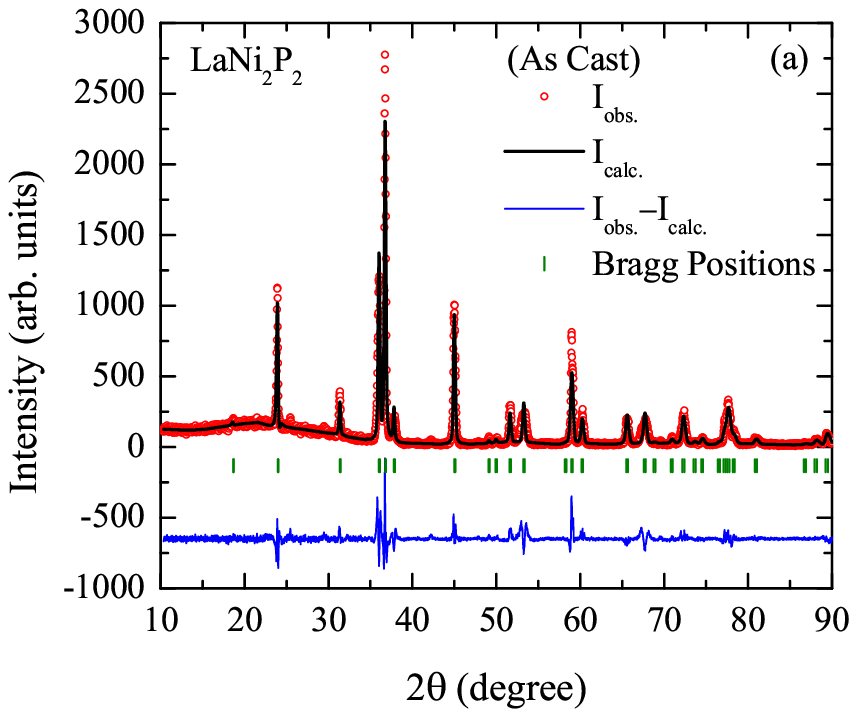}
	\includegraphics[width=3.3in]{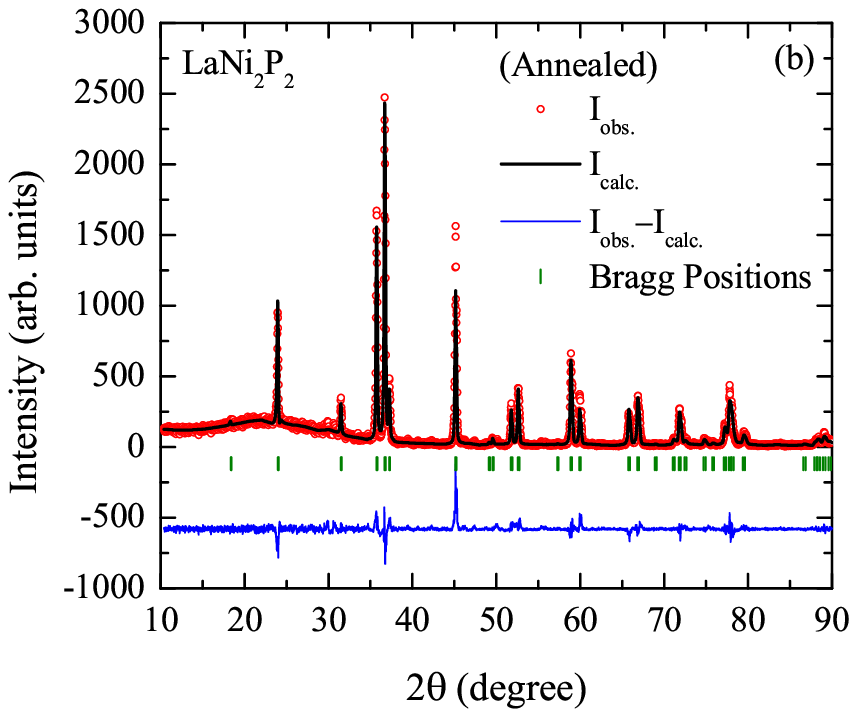}
	\caption{(Color online) Room temperature powder XRD pattern (red open circles) of LaNi$_2$P$_2$, Rietfeld refinement fit (solid black line), difference profile (lower solid blue line), and positions of Bragg peaks (vertical bars). The two panels show the XRD pattern obtained (a) before and (b) after annealing the sample.}
 	\label{fig:LaNi2P2_XRD}
\end{figure}

\begin{figure}
	\includegraphics[width=3.3in]{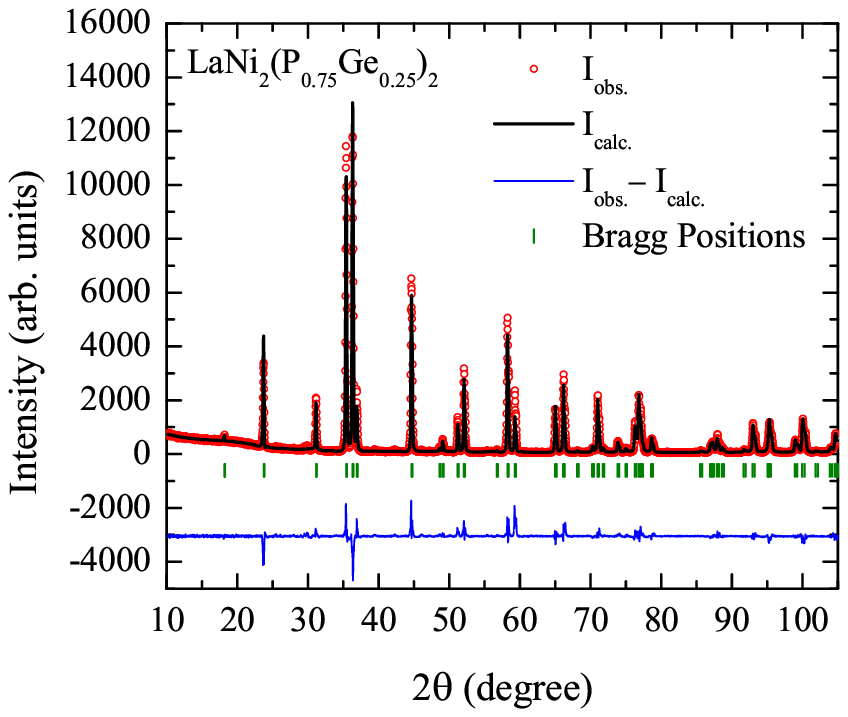}
	\caption{(Color online) Room temperature powder XRD pattern (red circles) of LaNi$_2$(P$_{0.75}$Ge$_{0.25}$)$_2$, Rietfeld refinement fit (solid black line), difference profile (lower solid blue line), and positions of Bragg peaks (vertical bars).}
	\label{fig:LaNi2P1.5Ge0.5_XRD}
\end{figure}

\begin{figure}
	\includegraphics[width=3.3in]{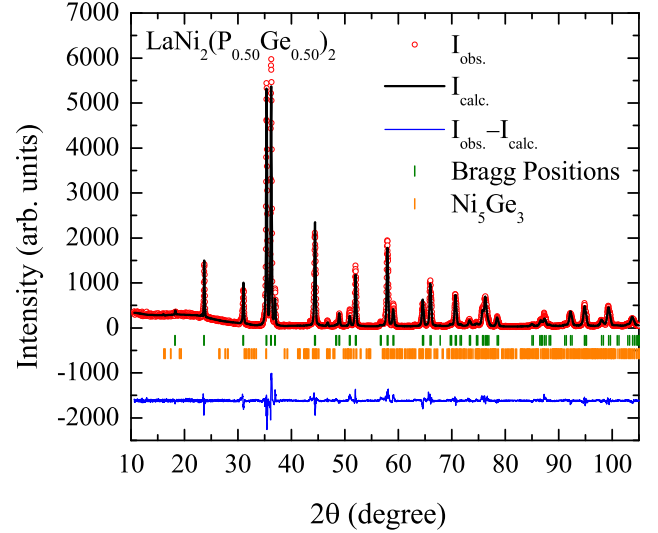}
	\caption{(Color online) Room temperature powder XRD pattern (red circles) of LaNi$_2$(P$_{0.50}$Ge$_{0.50}$)$_2$, multi-phase Rietfeld refinement fit (solid black line), difference profile (lower solid blue line), and positions of Bragg peaks (vertical bars; upper: LaNi$_2$(P$_{0.50}$Ge$_{0.50}$)$_2$, lower: Ni$_5$Ge$_3$). The refinement reveals that the phase composition is 99.6~mol\% LaNi$_2$(P$_{0.50}$Ge$_{0.50}$)$_2$ and 0.4~mol\% Ni$_5$Ge$_3$.}
	\label{fig:LaNi2PGe_XRD}
\end{figure}

\begin{figure}
	\includegraphics[width=3.3in]{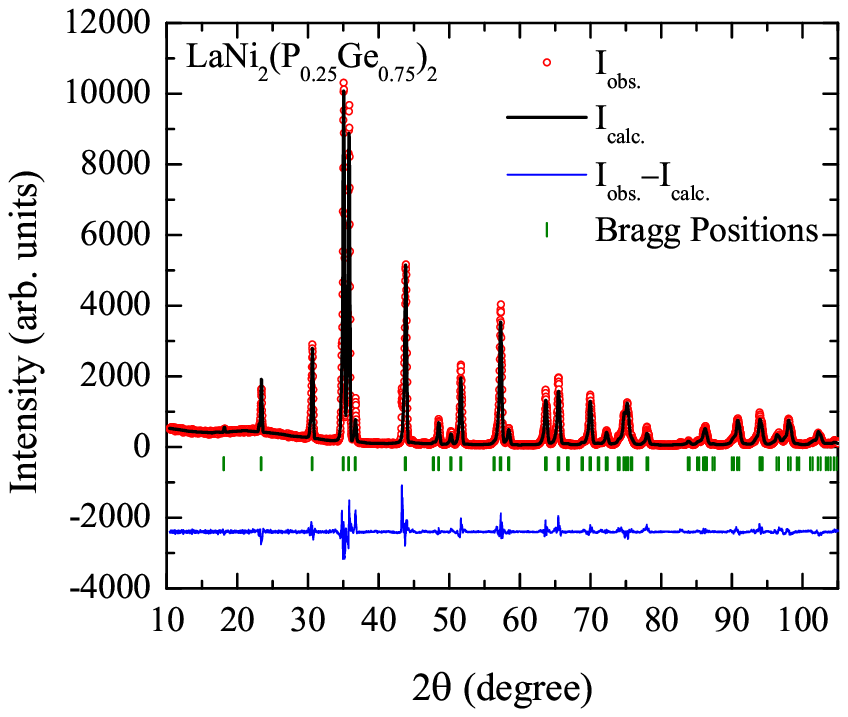}
	\caption{(Color online) Room temperature powder XRD pattern (red circles) of LaNi$_2$(P$_{0.25}$Ge$_{0.75}$)$_2$, Rietfeld refinement fit (solid black line), difference profile (lower solid blue line), and positions of Bragg peaks (vertical bars).}
	\label{fig:LaNi2P0.5Ge1.5_XRD}
\end{figure}

\begin{figure}
	\includegraphics[width=3.3in]{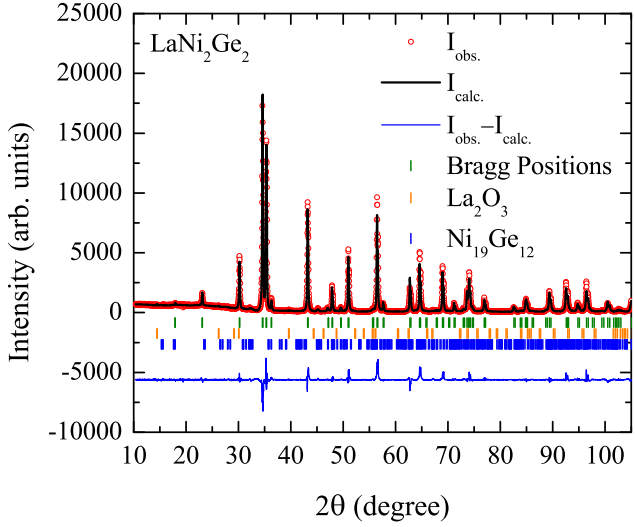}
	\caption{(Color online) Room temperature powder XRD pattern (red circles) of LaNi$_2$Ge$_2$, multi-phase Rietfeld refinement fit (solid black line), difference profile (lower solid blue line), and positions of Bragg peaks (vertical bars; upper: LaNi$_2$Ge$_2$, middle: La$_2$O$_3$, lower: Ni$_{19}$Ge$_{12}$). The refinement indicates that the phase composition of the sample is 93.9~mol\% LaNi$_2$Ge$_2$, 5.7~mol\% La$_2$O$_3$, and 0.4~mol\% Ni$_{19}$Ge$_{12}$.}
	\label{fig:LaNi2Ge2_XRD}
\end{figure}

In this Appendix the Rietveld refinement Figs.~\ref{fig:LaNi2P2_XRD}--\ref{fig:LaNi2Ge2_XRD} referred to in Sec.~\ref{Structure} are shown.

\clearpage

\section{\label{Sec:MHPlots} Magnetization versus Field Isotherms}

\begin{figure}
	\includegraphics[width=3.3in]{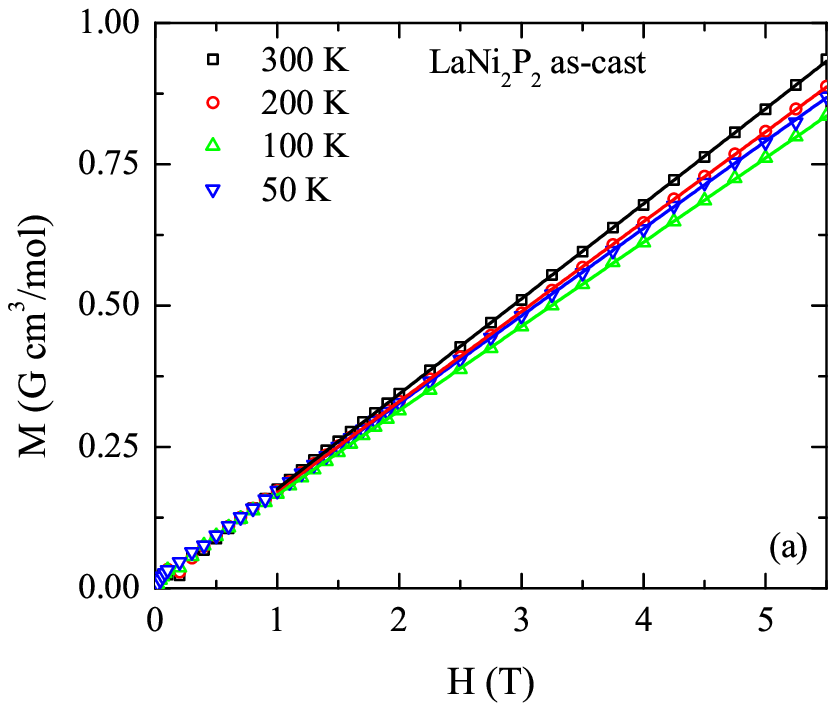}
	\includegraphics[width=3.3in]{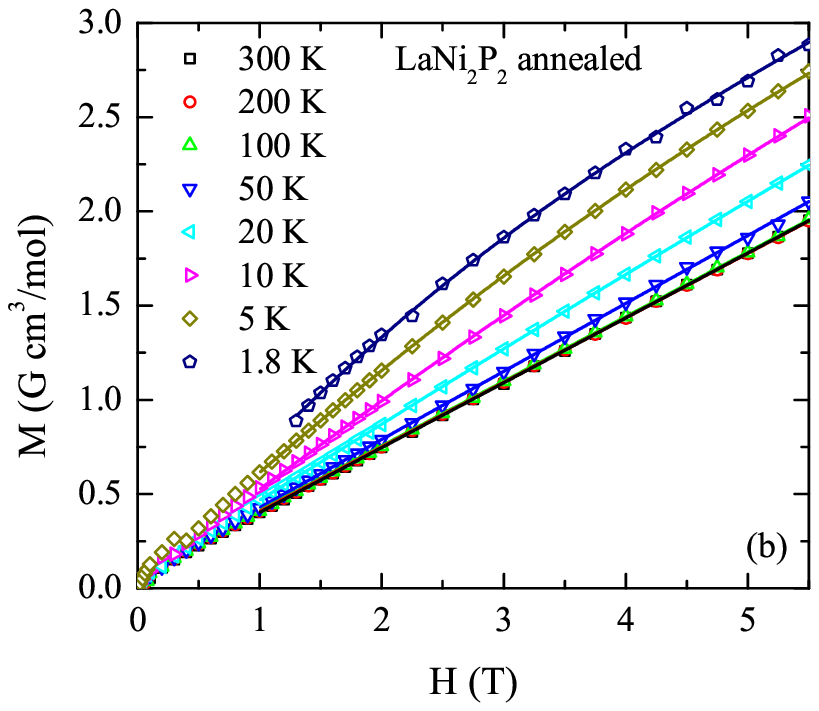}
 	\caption{(Color online) Magnetization $M$ versus applied field $H$ isotherms (symbols) at the listed temperatures for LaNi$_2$P$_2$ (a) before and (b) after annealing. The solid curves of the corresponding colors are fits by Eq. (\ref{eq:MH_fit}).}
	\label{fig:LaNi2P2_ascast_M_vs_H}
\end{figure}

\begin{figure}
	\includegraphics[width=3.3in]{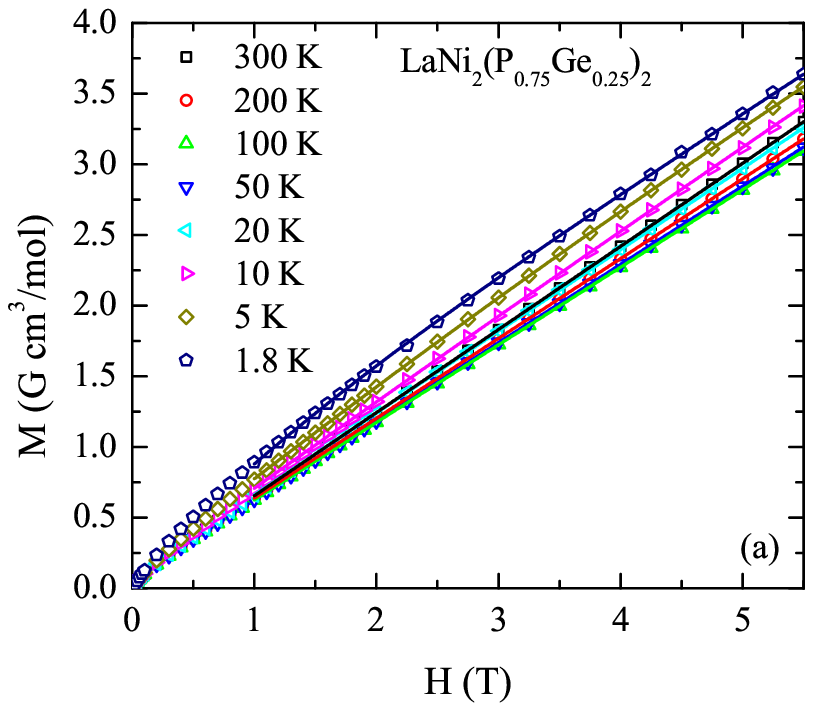}
	\includegraphics[width=3.3in]{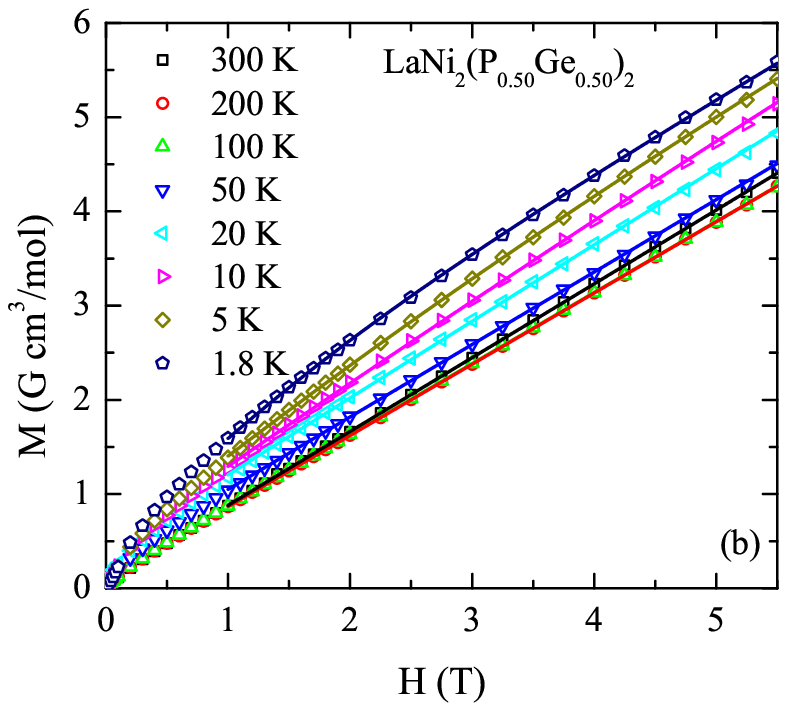}
	\includegraphics[width=3.3in]{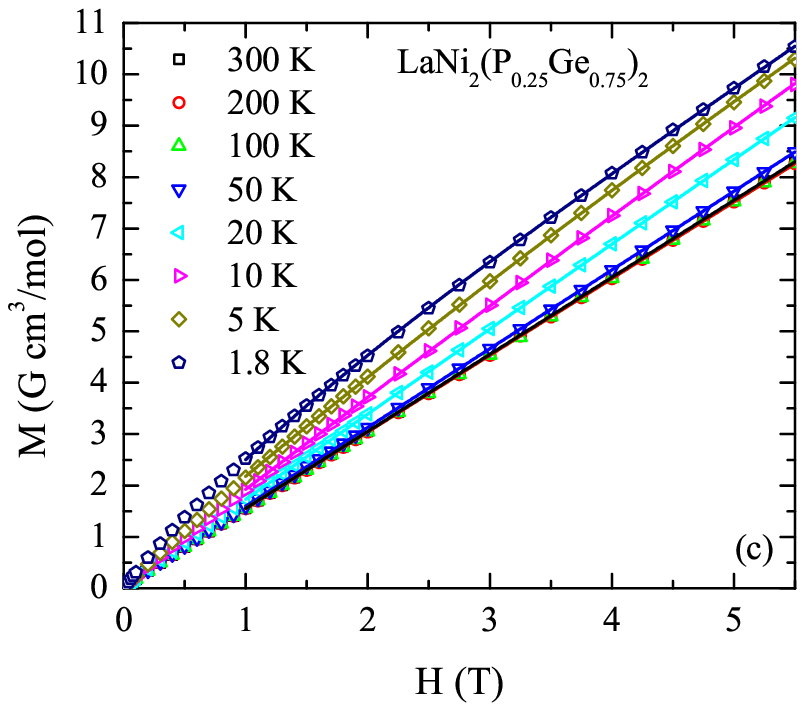}
 	\caption{(Color online) Magnetization $M$ versus applied magnetic field $H$ isotherms (symbols)  at the listed temperatures for LaNi$_2$(Ge$_{1-x}$P$_x$)$_2$ samples with (a) $x=0.25$, (b) $x=0.50$, and (c) $x=0.75$. In each figure, the solid curves of the corresponding colors are fits by Eq. (\ref{eq:MH_fit}).}
	\label{fig:LaNi2P0.5Ge1.5_M_vs_H}
\end{figure}

\begin{figure}
	\includegraphics[width=3.3in]{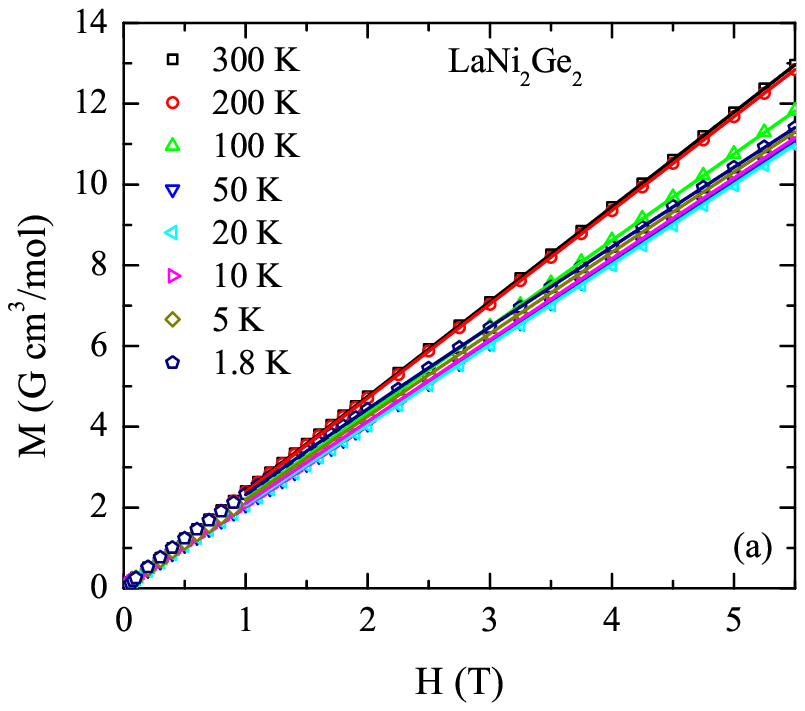}
	\includegraphics[width=3.3in]{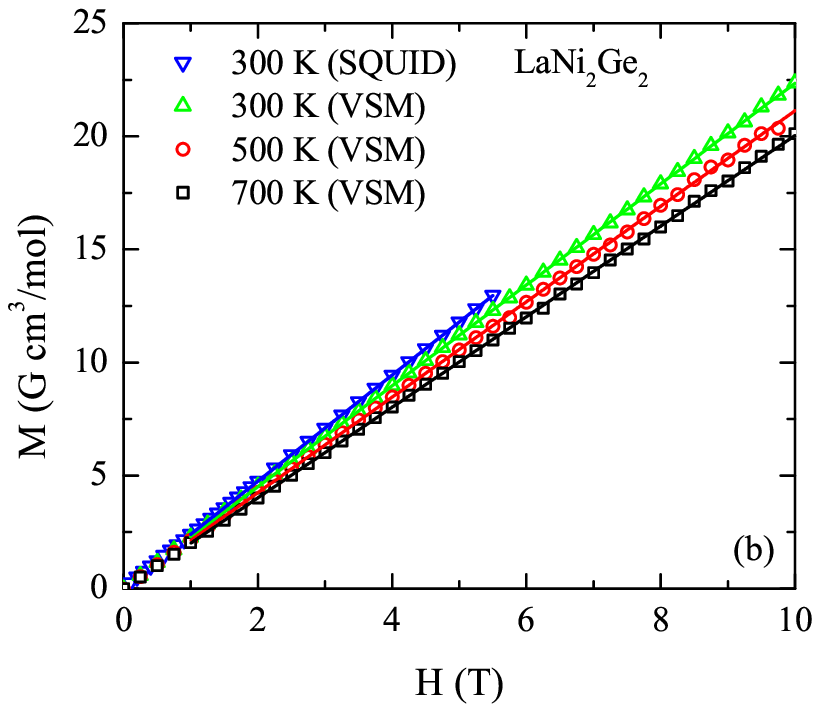}
 	\caption{(Color online) Magnetization $M$ versus applied field $H$ isotherms (symbols) at the listed temperatures for LaNi$_2$Ge$_2$ (a) using a SQUID magnetometer and (b) using a VSM magnetometer. The solid curves of the corresponding colors are fits by Eq.~(\ref{eq:MH_fit}).}
	\label{fig:LaNi2Ge2_M_vs_H}
\end{figure}

\begin{figure}
	\includegraphics[width=3.3in]{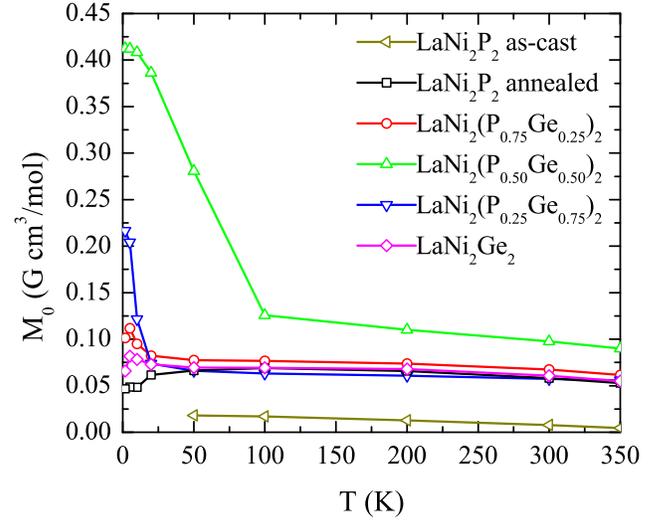}
 	\caption{(Color online) Saturation magnetization of ferromagnetic impurities $M_0$ in the LaNi$_2$(Ge$_{1-x}$P$_x$)$_2$ samples versus temperature $T$\@. The data at 350~K are extrapolated. The data for $x=0.5$ suggest the presence of a ferromagnetic impurity phase with a Curie temperature of $\approx 100$~K\@.  The solid lines are guides to the eye.}
	\label{fig:M0_vs_T}
\end{figure}

In this Appendix the $M(H)$ isotherms in Figs.~\ref{fig:LaNi2P2_ascast_M_vs_H}--\ref{fig:LaNi2Ge2_M_vs_H} and the dependence of the ferromagnetic impurity saturation magnetization on temperature in Fig.~\ref{fig:M0_vs_T} that are discussed in Sec.~\ref{Magnetic} are shown.

\end{document}